\PassOptionsToPackage{unicode}{hyperref}
\PassOptionsToPackage{hyphens}{url}
\PassOptionsToPackage{dvipsnames,svgnames,x11names}{xcolor}
\documentclass[
]{article}
\usepackage{xcolor}
\usepackage{amsmath,amssymb}
\usepackage{iftex}
\ifPDFTeX
  \usepackage[T1]{fontenc}
  \usepackage[utf8]{inputenc}
  \usepackage{textcomp} 
\else 
  \usepackage{unicode-math} 
  \defaultfontfeatures{Scale=MatchLowercase}
  \defaultfontfeatures[\rmfamily]{Ligatures=TeX,Scale=1}
\fi
\usepackage{lmodern}
\ifPDFTeX\else
\fi
\IfFileExists{upquote.sty}{\usepackage{upquote}}{}
\IfFileExists{microtype.sty}{
  \usepackage[]{microtype}
  \UseMicrotypeSet[protrusion]{basicmath} 
}{}
\makeatletter
\@ifundefined{KOMAClassName}{
  \IfFileExists{parskip.sty}{%
    \usepackage{parskip}
  }{
    \setlength{\parindent}{0pt}
    \setlength{\parskip}{6pt plus 2pt minus 1pt}}
}{
  \KOMAoptions{parskip=half}}
\makeatother
\makeatletter
\ifx\paragraph\undefined\else
  \let\oldparagraph\paragraph
  \renewcommand{\paragraph}{
    \@ifstar
      \xxxParagraphStar
      \xxxParagraphNoStar
  }
  \newcommand{\xxxParagraphStar}[1]{\oldparagraph*{#1}\mbox{}}
  \newcommand{\xxxParagraphNoStar}[1]{\oldparagraph{#1}\mbox{}}
\fi
\ifx\subparagraph\undefined\else
  \let\oldsubparagraph\subparagraph
  \renewcommand{\subparagraph}{
    \@ifstar
      \xxxSubParagraphStar
      \xxxSubParagraphNoStar
  }
  \newcommand{\xxxSubParagraphStar}[1]{\oldsubparagraph*{#1}\mbox{}}
  \newcommand{\xxxSubParagraphNoStar}[1]{\oldsubparagraph{#1}\mbox{}}
\fi
\makeatother

\usepackage{color}
\usepackage{fancyvrb}

\DefineVerbatimEnvironment{Highlighting}{Verbatim}{commandchars=\\\{\}}
\usepackage{framed}
\definecolor{shadecolor}{RGB}{241,243,245}
\newenvironment{Shaded}{\begin{snugshade}}{\end{snugshade}}

\newcommand{\AttributeTok}[1]{\textcolor[rgb]{0.40,0.45,0.13}{#1}}

\newcommand{\CommentTok}[1]{\textcolor[rgb]{0.37,0.37,0.37}{#1}}

\newcommand{\ConstantTok}[1]{\textcolor[rgb]{0.56,0.35,0.01}{#1}}
\newcommand{\ControlFlowTok}[1]{\textcolor[rgb]{0.00,0.23,0.31}{\textbf{#1}}}

\newcommand{\DecValTok}[1]{\textcolor[rgb]{0.68,0.00,0.00}{#1}}

\newcommand{\ErrorTok}[1]{\textcolor[rgb]{0.68,0.00,0.00}{#1}}

\newcommand{\FloatTok}[1]{\textcolor[rgb]{0.68,0.00,0.00}{#1}}
\newcommand{\FunctionTok}[1]{\textcolor[rgb]{0.28,0.35,0.67}{#1}}

\newcommand{\NormalTok}[1]{\textcolor[rgb]{0.00,0.23,0.31}{#1}}

\newcommand{\OtherTok}[1]{\textcolor[rgb]{0.00,0.23,0.31}{#1}}

\newcommand{\SpecialCharTok}[1]{\textcolor[rgb]{0.37,0.37,0.37}{#1}}

\newcommand{\StringTok}[1]{\textcolor[rgb]{0.13,0.47,0.30}{#1}}

\usepackage{longtable,booktabs,array}
\usepackage{calc} 
\usepackage{etoolbox}
\makeatletter
\patchcmd\longtable{\par}{\if@noskipsec\mbox{}\fi\par}{}{}
\makeatother
\IfFileExists{footnotehyper.sty}{\usepackage{footnotehyper}}{\usepackage{footnote}}
\makesavenoteenv{longtable}
\usepackage{graphicx}
\makeatletter
\newsavebox\pandoc@box
\newcommand*\pandocbounded[1]{
  \sbox\pandoc@box{#1}%
  \Gscale@div\@tempa{\textheight}{\dimexpr\ht\pandoc@box+\dp\pandoc@box\relax}%
  \Gscale@div\@tempb{\linewidth}{\wd\pandoc@box}%
  \ifdim\@tempb\p@<\@tempa\p@\let\@tempa\@tempb\fi
  \ifdim\@tempa\p@<\p@\scalebox{\@tempa}{\usebox\pandoc@box}%
  \else\usebox{\pandoc@box}%
  \fi%
}
\def\fps@figure{htbp}
\makeatother

\providecommand{\tightlist}{%
  \setlength{\itemsep}{0pt}\setlength{\parskip}{0pt}}

\usepackage[]{natbib}
\usepackage{arxiv}
\usepackage{orcidlink}
\usepackage{amsmath}
\usepackage[T1]{fontenc}
\makeatletter
\@ifpackageloaded{caption}{}{\usepackage{caption}}
\AtBeginDocument{%
\ifdefined\contentsname
  \renewcommand*\contentsname{Table of contents}
\else
  \newcommand\contentsname{Table of contents}
\fi
\ifdefined\listfigurename
  \renewcommand*\listfigurename{List of Figures}
\else
  \newcommand\listfigurename{List of Figures}
\fi
\ifdefined\listtablename
  \renewcommand*\listtablename{List of Tables}
\else
  \newcommand\listtablename{List of Tables}
\fi
\ifdefined\figurename
  \renewcommand*\figurename{Figure}
\else
  \newcommand\figurename{Figure}
\fi
\ifdefined\tablename
  \renewcommand*\tablename{Table}
\else
  \newcommand\tablename{Table}
\fi
}
\@ifpackageloaded{float}{}{\usepackage{float}}
\floatstyle{ruled}
\@ifundefined{c@chapter}{\newfloat{codelisting}{h}{lop}}{\newfloat{codelisting}{h}{lop}[chapter]}
\floatname{codelisting}{Listing}

\makeatother
\makeatletter
\@ifpackageloaded{caption}{}{\usepackage{caption}}
\@ifpackageloaded{subcaption}{}{\usepackage{subcaption}}
\makeatother
\usepackage{bookmark}
\IfFileExists{xurl.sty}{\usepackage{xurl}}{} 
\hypersetup{
  pdftitle={rush: Scalable Asynchronous Distributed Computing via Shared State in R},
  pdfauthor={Marc Becker; Bernd Bischl},
  colorlinks=true,
  linkcolor={blue},
  filecolor={Maroon},
  citecolor={Blue},
  urlcolor={Blue},
  pdfcreator={LaTeX via pandoc}}

\renewcommand{\today}{2026-09-02}
\newcommand{\runninghead}{A Preprint }
\renewcommand{\runninghead}{A Preprint }
\title{rush: Scalable Asynchronous Distributed Computing via Shared
State in R}
\def\asep{\\\\\\ } 
\author{\textbf{Marc Becker}~\orcidlink{0000-0002-8115-0400}\\Department
of Statistics\\Ludwig-Maximilians-Universität München\\Munich\\\\Munich
Center for Machine Learning
(MCML)\\Munich\\\href{mailto:marc.becker@stat.uni-muenchen.de}{marc.becker@stat.uni-muenchen.de}\asep\textbf{Bernd
Bischl}~\orcidlink{0000-0001-6002-6980}\\Department of
Statistics\\Ludwig-Maximilians-Universität München\\Munich\\\\Munich
Center for Machine Learning
(MCML)\\Munich\\\href{mailto:bernd.bischl@stat.uni-muenchen.de}{bernd.bischl@stat.uni-muenchen.de}}
\date{2026-09-02}
\begin{document}
\maketitle
\begin{abstract}
Many algorithms in statistics and machine learning can be parallelized
in an asynchronous manner where workers need to communicate through
shared state rather than execute independent tasks dispatched by a
central controller. Especially in modern hyperparameter optimization and
parallel black-box optimization with expensive objectives, this
decentralized approach has become widespread, and several Python
frameworks adopt it (e.g., Optuna, DeepHyper, and Hyperopt). However,
all popular R packages for parallel computing follow a centralized
controller-worker architecture that does not support this pattern. We
present rush, an R package that provides a shared-state coordination
layer for asynchronously parallelized iterative algorithms. rush uses a
Redis database as a shared key-value store: workers read and write task
data through the database and independently execute their own loops. The
package provides a high-level API for managing tasks and their
lifecycle, featuring sub-millisecond per-task overhead, robust error
handling with automatic detection of lost workers, and efficient
caching. rush optionally integrates with the mlr3 ecosystem, powering
asynchronous optimization in the bbotk and mlr3tuning packages. We
demonstrate the practical utility of rush by implementing asynchronous
decentralized Bayesian optimization (ADBO) and benchmarking it on
hyperparameter optimization of LightGBM across four datasets using 448
workers.
\end{abstract}

\section{Introduction}\label{sec-introduction}

Many iterative algorithms in statistics and machine learning share a
common computational structure: multiple workers operate concurrently,
each maintaining local state and autonomously deciding what to compute
next, while coordinating through a shared pool of (intermediate)
results. This computational pattern has long-standing roots in the
\emph{blackboard architecture} \citep{nii1986} and the \emph{tuple
space} model \citep{gelernter1985}, coordination paradigms in which
autonomous processes communicate through a shared associative memory.

Frameworks for parallel computation have become extremely popular in
modern machine learning in the context of expensive black-box
optimization and hyperparameter optimization \citep[HPO,][]{bischl2023}.
Here, algorithmic approaches span a wide spectrum: from synchronous
batch methods \citep{ginsbourger2010}, through asynchronous but
centralized schedulers such as ASHA \citep{li2020}\footnote{ASHA is
  algorithmically decentralized, but its common implementation in Ray
  Tune uses a central scheduler.} and Vizier \citep{golovin2017}, to
fully decentralized approaches such as asynchronous decentralized
Bayesian optimization \citep[ADBO,][]{egele2023}, where each worker
autonomously decides what to compute next.

\begin{figure}

\centering{

\includegraphics[width=1\linewidth,height=\textheight,keepaspectratio,alt={Diagrams of a centralized network with a main process connected to workers and a decentralized network with workers connected to a shared Redis database.}]{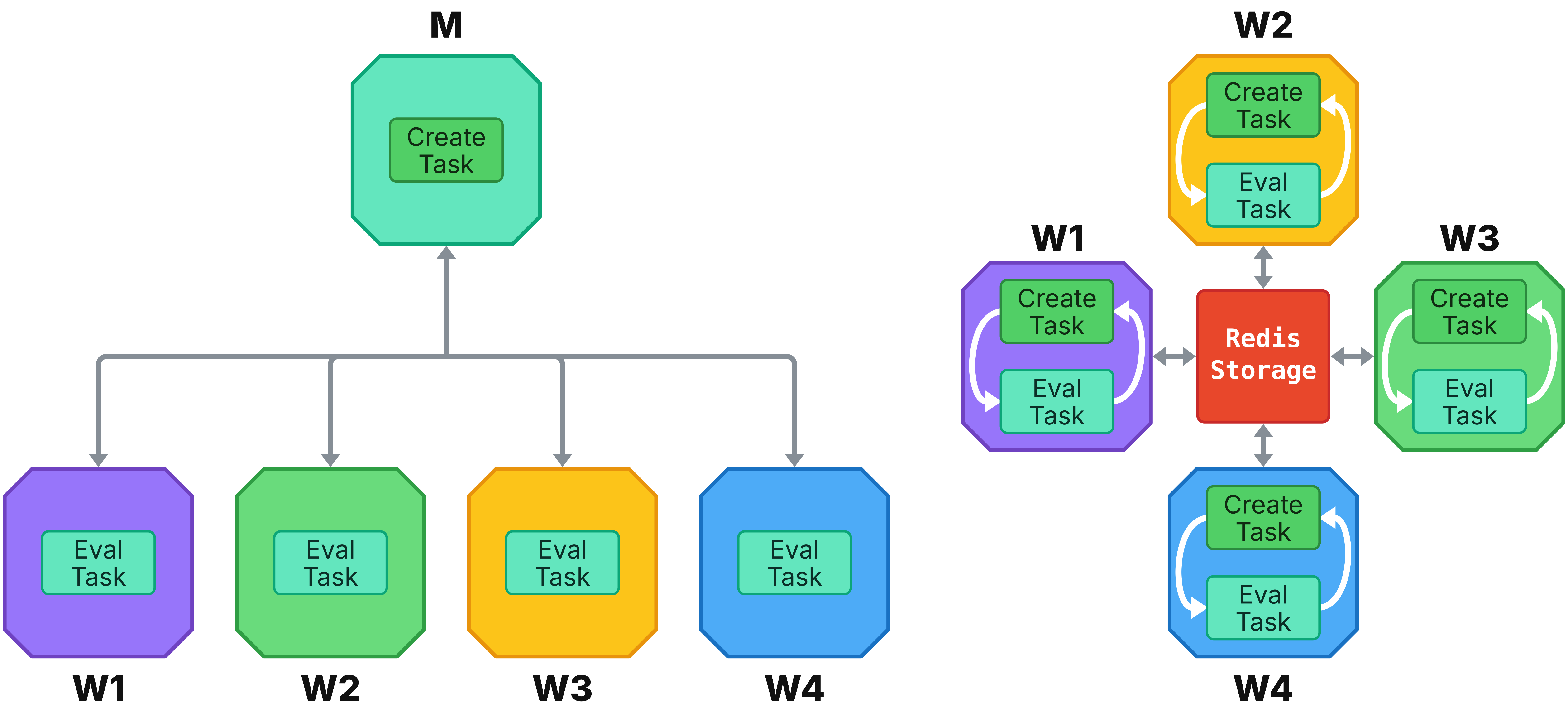}

}

\caption{\label{fig-rush-network}Centralized (left) and decentralized
(right) networks. Octagons denote processes, including the main process
\texttt{M} and workers \texttt{W}. In the centralized network,
\texttt{M} proposes tasks to the workers and processes the returned
computations. In the decentralized network, each worker \texttt{W}
independently proposes tasks, evaluates them, and exchanges task
information via a shared \texttt{Redis} database (red rectangle).}

\end{figure}

In contrast to these decentralized approaches, existing R packages for
parallel computing follow a fundamentally different design. Packages
such as \texttt{parallel}, \texttt{future} \citep{bengtsson2021},
\texttt{mirai} \citep{gao2025}, \texttt{batchtools} \citep{lang2017},
and \texttt{crew} \citep{landau2026} adopt a controller-worker
architecture (Figure~\ref{fig-rush-network}, left): a central process
proposes tasks, dispatches them to workers, and collects results. These
frameworks excel at parallelizing independent computations but do not
support algorithms in which workers coordinate through shared state and
autonomously decide what to compute next. Beyond expressiveness --- some
algorithms are simply more natural when each worker has the coordination
logic locally --- the decentralized design also has computational
advantages. Planning and generating further downstream tasks can
themselves incur non-negligible cost, and moving this work from a
central node to the workers allows it to run in parallel and enables
more efficient task spawning, especially when many workers are
available.

To fill this gap, we present \texttt{rush}, an R package that provides a
shared-state coordination layer for iterative algorithms. \texttt{rush}
uses a \texttt{Redis} database as its shared store: workers communicate
by reading and writing task data---inputs, outputs, and computational
states---through the database, each independently executing its own loop
(Figure~\ref{fig-rush-network}, right). The package provides a
high-level API for managing tasks and their lifecycle, featuring
sub-millisecond per-task overhead, robust error handling with automatic
detection of lost workers, and an efficient caching mechanism that
minimizes database operations. \texttt{rush} integrates with the
\texttt{mlr3} \citep{lang2019} ecosystem, powering asynchronous
optimization in the \texttt{bbotk} \citep{becker2026b} and
\texttt{mlr3tuning} \citep{becker2025a} packages.

While \texttt{rush} is designed as a general coordination layer and can
be used for a wide range of algorithms, we demonstrate its utility
through a case study in black-box optimization --- which was also our
primary motivation for developing \texttt{rush}. We implement ADBO
\citep{egele2023} on \texttt{rush} and benchmark it across four datasets
using 448 workers. Our experiments show that ADBO on \texttt{rush}
achieves a CPU utilization above 93\% in every run, compared to
single-digit utilization for centralized approaches on short-running
tasks.

\subsection{Related work}\label{sec-related-work}

The widespread adoption of multi-core processors in the 2000s created a
growing need to utilize these resources effectively for computational
tasks in R. The first packages to address this need were \texttt{snow}
\citep{rossini2007} and \texttt{multicore}. With R version 2.14.0
(released in 2011), parallel computing capabilities were integrated into
the base R system through the \texttt{parallel} package. The
\texttt{parallel} package provides \texttt{mclapply()} and
\texttt{parLapply()} as parallel versions of \texttt{lapply()} for
multicore and cluster computing, respectively. Although widely used,
both functions block the R session until all tasks have completed,
preventing retrieval of partial results. Moreover, their static task
allocation can lead to poor load balancing when tasks have heterogeneous
runtimes.

The \texttt{future} package \citep{bengtsson2021} introduced a unified
parallel computing interface in R, supporting various backends such as
\texttt{multisession}, \texttt{multicore}, and \texttt{callr}, with the
\texttt{future.apply} package providing parallel versions of the
\texttt{*apply()} family of functions. However, the
\texttt{multisession} and \texttt{cluster} backends are limited to 125
workers. Furthermore, per-task overhead --- from process spawning
(\texttt{multicore}, \texttt{callr}) or serialization across persistent
workers (\texttt{multisession}, \texttt{cluster}) --- can be significant
for short-running tasks.

The \texttt{mirai} package \citep{gao2025} evaluates an R expression
asynchronously in a parallel process, either locally or distributed
across the network. The package has a strong focus on throughput, low
overhead, and modern networking, eliminating many drawbacks of the older
approaches. \texttt{mirai} provides a comprehensive daemon management
interface for launching and configuring persistent daemons across local
and remote systems (including SSH-based deployment and HPC schedulers).
\texttt{rush} leverages this mechanism to start workers on remote
machines.

The \texttt{batchtools} package \citep{lang2017} facilitates the
execution of long-running tasks and large-scale computer experiments on
HPC clusters. Because communication between the main process and workers
relies entirely on the file system, the package incurs substantial
per-task overhead from both job scheduling and file I/O, making it less
suitable for high-throughput workloads with many short-running tasks.
The more recent \texttt{crew} package \citep{landau2026} targets a
similar use case, supporting long-running tasks in distributed systems
ranging from traditional HPC clusters to cloud computing platforms, but
likewise depends on a centralized controller for task coordination.

The parallel and distributed computing ecosystem in R is complemented by
database connectivity layers that enable shared state and coordination.
For relational database management systems, \texttt{DBI}
\citep{rsigdb2024} defines a standardized interface that separates a
common frontend from database-specific backends. For \texttt{Redis},
\texttt{redux} provides R bindings to the \texttt{Redis} API and exposes
the command set through a low-level and a high-level interface.

The \texttt{rrq} \citep{fitzjohn2026} package is a task queue system for
R using \texttt{Redis} as a coordination backend. It addresses the
limitations of other packages by providing a non-blocking interface to
parallel computing and keeping the overhead per task low. The package
supports independent task queues, priority levels within queues, and
inter-task dependencies. \texttt{rrq} is implemented on top of
\texttt{redux}, which provides the underlying \texttt{Redis}
connectivity and command interface. Its error-handling mechanism,
particularly its heartbeat approach, directly informed the design of the
corresponding mechanism in \texttt{rush}.

Python has a rich ecosystem for parallel and distributed computing.
\texttt{Dask} \citep{rocklin2015} enables parallel and distributed
execution via a dynamic task scheduler, and \texttt{Ray}
\citep{moritz2018} is a general-purpose distributed computing framework.
For HPO specifically, many frameworks adopt a centralized architecture
in which a central scheduler dispatches configurations to workers:
\texttt{Ray\ Tune} \citep{liaw2018}, \texttt{Syne\ Tune}
\citep{salinas2022}, \texttt{Google\ Vizier} \citep{golovin2017},
\texttt{Ax} \citep{olson2025}, \texttt{Dragonfly} \citep{kandasamy2020},
and \texttt{Keras\ Tuner} \citep{omalley2019}. In contrast, frameworks
such as \texttt{Optuna} \citep{akiba2019}, \texttt{DeepHyper}
\citep{egele2025}, \texttt{Hyperopt} \citep{bergstra2013a}, and
\texttt{Orion} \citep{bouthillier2023} support decentralized operation
by coordinating workers through a shared database (SQL,
\texttt{MongoDB}, or \texttt{Redis}), enabling each worker to
independently propose and evaluate configurations.

In summary, the existing R ecosystem for parallel and distributed
computing is rich but uniformly follows a controller-worker
architecture. Packages such as \texttt{parallel}, \texttt{future}, and
\texttt{crew} centralize task proposal and result collection in a single
process, which can become a bottleneck at scale. While \texttt{rrq} and
\texttt{mirai} reduce per-task overhead, neither supports decentralized
optimization loops in which workers independently propose and evaluate
tasks. At least in the domains of expensive black-box optimization and
HPO, Python offers frameworks such as \texttt{DeepHyper} and
\texttt{Optuna} that support decentralized workflows, but no comparable
solution exists in R. \texttt{rush} addresses this gap by combining a
decentralized architecture with low-overhead \texttt{Redis}-based
communication.

\newpage

\section{Architecture and API of rush}\label{sec-general-structure}

\begin{figure}

\centering{

\includegraphics[width=0.6\linewidth,height=\textheight,keepaspectratio,alt={Diagram of the communication flow between a worker and the Redis database in a rush network.}]{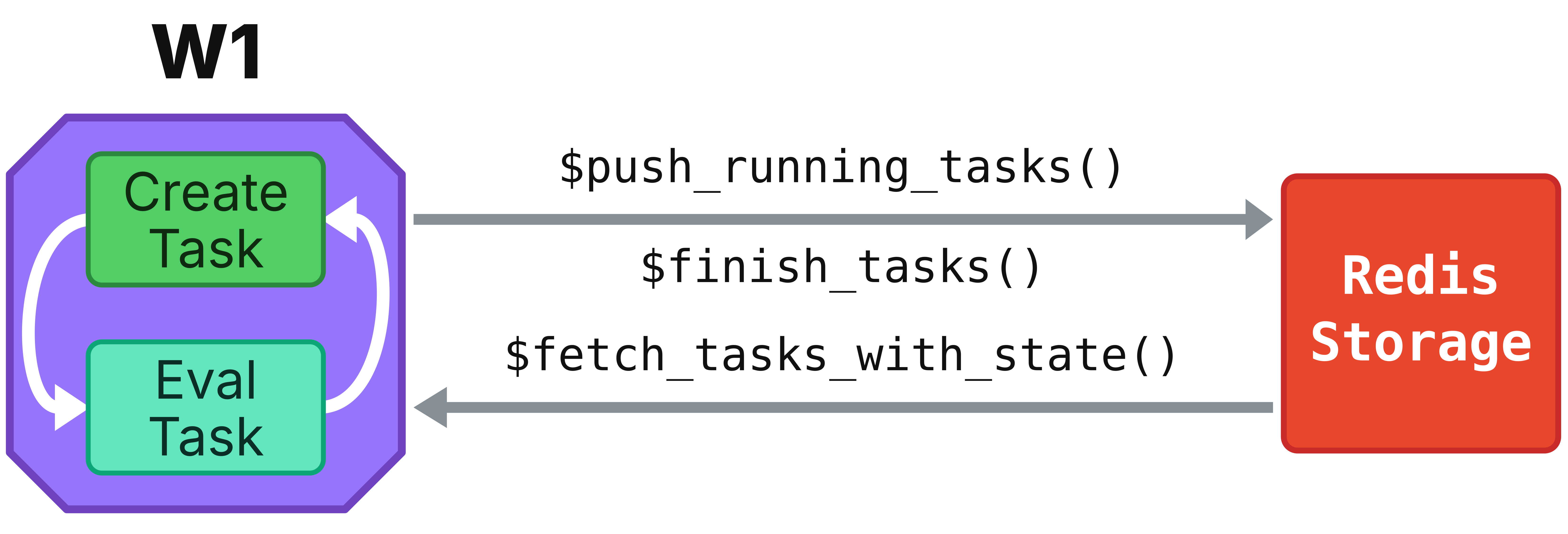}

}

\caption{\label{fig-worker-communication}The communication flow between
a worker and the \texttt{Redis} database in a \texttt{rush} network. The
octagon represents a worker and the rectangle represents the
\texttt{Redis} database. In each iteration, the worker reads shared
state via \texttt{\$fetch\_tasks\_with\_state()}, registers a new task
via \texttt{\$push\_running\_tasks()}, computes the result, and writes
it back via \texttt{\$finish\_tasks()}.}

\end{figure}

This section introduces the architecture and core API of \texttt{rush}.
The package is released on CRAN and its development version is hosted at
https://github.com/mlr-org/rush. A \texttt{rush} network consists of
multiple workers that communicate via a shared \texttt{Redis} database.
A worker can read shared state from the database, compute new (partial)
results, and potentially write them back into the shared database, as
illustrated in Figure~\ref{fig-worker-communication}; see
Figure~\ref{fig-rush-network} (right) for the full network topology.

\subsection{Manager}\label{sec-manager}

The \texttt{Rush} manager class is responsible for creating, monitoring,
and stopping a network. It is initialized using the \texttt{rsh()}
function, which requires a network identifier and a \texttt{config}
argument that specifies how to connect to the \texttt{Redis} database
via the \texttt{redux} package. A running \texttt{Redis} server must be
started separately beforehand.\footnote{Installation instructions for
  all platforms are available at
  https://redis.io/docs/latest/operate/oss\_and\_stack/install/install-redis/.}

\begin{Shaded}
\begin{Highlighting}[]
\FunctionTok{library}\NormalTok{(rush)}

\NormalTok{config }\OtherTok{=}\NormalTok{ redux}\SpecialCharTok{::}\FunctionTok{redis\_config}\NormalTok{()}

\NormalTok{rush }\OtherTok{=} \FunctionTok{rsh}\NormalTok{(}\AttributeTok{network =} \StringTok{"demo{-}network"}\NormalTok{, }\AttributeTok{config =}\NormalTok{ config)}
\end{Highlighting}
\end{Shaded}

Workers are started using the \texttt{\$start\_workers()} method, which
accepts the worker loop (see Section~\ref{sec-worker-loop}) and the
number of workers as arguments. Any additional named arguments are
forwarded to the worker loop function. Workers run on \texttt{mirai}
daemons \citep{gao2025}, which are launched via
\texttt{mirai::daemons()}. Calling \texttt{mirai::daemons(n)} opens a
listener socket on the host, spawns a lightweight dispatcher implemented
in C, and launches \texttt{n} persistent R processes that dial into the
dispatcher and enter a receive--evaluate--return loop. Note that while
\texttt{rush} still uses this mechanism to set up workers, most of the
communication is actually facilitated via the \texttt{Redis} database,
e.g., worker loop arguments, task inputs, outputs, and states. Daemons
can be spawned on the local machine, on remote nodes via SSH, or on HPC
clusters via job schedulers such as SLURM or SGE --- in all cases, the
daemon process dials back to the host and the same \texttt{rush} worker
loop runs transparently. Alternatively, workers can be started locally
via the \texttt{processx} package using
\texttt{\$start\_local\_workers()}, or a standalone script can be
generated with \texttt{\$worker\_script()} for manual deployment; the
only requirement is that the worker can connect to the \texttt{Redis}
database.

\begin{Shaded}
\begin{Highlighting}[]
\NormalTok{mirai}\SpecialCharTok{::}\FunctionTok{daemons}\NormalTok{(}\DecValTok{4}\NormalTok{)}

\NormalTok{rush}\SpecialCharTok{$}\FunctionTok{start\_workers}\NormalTok{(}
  \AttributeTok{worker\_loop =}\NormalTok{ worker\_loop,}
  \AttributeTok{n\_workers =} \DecValTok{4}\NormalTok{)}
\NormalTok{rush}\SpecialCharTok{$}\FunctionTok{wait\_for\_workers}\NormalTok{(}\DecValTok{4}\NormalTok{)}
\end{Highlighting}
\end{Shaded}

Since \texttt{\$start\_workers()} returns immediately,
\texttt{\$wait\_for\_workers(n)} blocks until \texttt{n} workers have
registered in \texttt{Redis}. Printing the \texttt{Rush} object reports
the number of running workers and a summary of task states across the
network (see Section~\ref{sec-tasks}).

\begin{Shaded}
\begin{Highlighting}[]
\NormalTok{rush}
\end{Highlighting}
\end{Shaded}

\begin{verbatim}

-- <Rush> --------------------------------------------------------------
* Running Workers: 4
* Queued Tasks: 0
* Running Tasks: 0
* Finished Tasks: 0
* Failed Tasks: 0
\end{verbatim}

The \texttt{\$worker\_info} active binding returns a \texttt{data.table}
\citep{barrett2026} with one row per worker, containing the worker
identifier, the process ID, the host on which it runs, whether a
heartbeat (see Section~\ref{sec-error-handling}) is active, and its
current state.

\begin{Shaded}
\begin{Highlighting}[]
\NormalTok{rush}\SpecialCharTok{$}\NormalTok{worker\_info}
\end{Highlighting}
\end{Shaded}

\begin{verbatim}
          worker_id   pid         hostname profile heartbeat   state
             <char> <int>           <char>  <char>    <lgcl>  <char>
1: pseudopolitic... 37025 ThinkPad-T14-...    <NA>     FALSE running
2: idiocratic_ca... 37028 ThinkPad-T14-...    <NA>     FALSE running
3: papery_sloth_... 37023 ThinkPad-T14-...    <NA>     FALSE running
4: precarnival_b... 37021 ThinkPad-T14-...    <NA>     FALSE running
\end{verbatim}

The \texttt{profile} column identifies each worker's compute profile,
which can be used to group workers by available resources, e.g., CPUs or
GPUs.

The workers can be stopped individually via
\texttt{\$stop\_workers(ids)} or all together using the
\texttt{\$reset()} method, which also resets the network and database.

\subsection{Worker loop}\label{sec-worker-loop}

A worker loop is a plain R function that receives a \texttt{RushWorker}
object for communicating with \texttt{Redis}.

\begin{Shaded}
\begin{Highlighting}[]
\NormalTok{worker\_loop }\OtherTok{=} \ControlFlowTok{function}\NormalTok{(rush, n\_evals) \{}
  \ControlFlowTok{while}\NormalTok{ (rush}\SpecialCharTok{$}\NormalTok{n\_finished\_tasks }\SpecialCharTok{\textless{}}\NormalTok{ n\_evals) \{}
\NormalTok{    archive }\OtherTok{=}\NormalTok{ rush}\SpecialCharTok{$}\FunctionTok{fetch\_tasks\_with\_state}\NormalTok{(}\AttributeTok{states =} \FunctionTok{c}\NormalTok{(}\StringTok{"running"}\NormalTok{, }\StringTok{"finished"}\NormalTok{))}
\NormalTok{    xs }\OtherTok{=} \FunctionTok{compute\_task\_inputs}\NormalTok{(archive)}
\NormalTok{    keys }\OtherTok{=}\NormalTok{ rush}\SpecialCharTok{$}\FunctionTok{push\_running\_tasks}\NormalTok{(}\AttributeTok{xss =} \FunctionTok{list}\NormalTok{(xs))}
\NormalTok{    ys }\OtherTok{=} \FunctionTok{compute\_task\_results}\NormalTok{(xs)}
\NormalTok{    rush}\SpecialCharTok{$}\FunctionTok{finish\_tasks}\NormalTok{(keys, }\AttributeTok{yss =} \FunctionTok{list}\NormalTok{(ys))}
\NormalTok{  \}}
\NormalTok{\}}
\end{Highlighting}
\end{Shaded}

The above shows an abstract template for an exemplary situation where
\texttt{rush} excels --- and which is already fairly close to the
asynchronous, decentralized HPO use case that we will cover in more
detail in Section~\ref{sec-case-study}. In the (common) example above,
we assume the following:

\begin{itemize}
\tightlist
\item
  The worker needs access to all available partial results to define and
  compute its next task.
\item
  The inputs for the task also have to be computed and thus incur
  non-negligible costs.
\item
  The computation of the task itself is time-consuming.
\end{itemize}

Inside the loop, the worker communicates with the shared \texttt{Redis}
database through the following core methods of the \texttt{RushWorker}
object:

\begin{itemize}
\tightlist
\item
  \texttt{\$push\_running\_tasks(xss)} - Creates one or more new tasks
  with inputs \texttt{xss} in the database, marks them as
  \texttt{"running"}, and returns their unique keys. Called before
  evaluation begins.
\item
  \texttt{\$fetch\_finished\_tasks()} - Retrieves all finished tasks
  from the database as a \texttt{data.table} (keys, inputs, outputs).
  Supports caching: previously fetched tasks are served from a local
  cache, and only new results are read from \texttt{Redis}.
\item
  \texttt{\$fetch\_tasks\_with\_state(states)} - More general variant
  that retrieves tasks in specified states (e.g., both
  \texttt{"running"} and \texttt{"finished"}), useful when the worker
  needs to account for evaluations still in progress.
\item
  \texttt{\$finish\_tasks(keys,\ yss)} - For tasks named by
  \texttt{keys}, writes corresponding results \texttt{yss} to the
  database and marks them as \texttt{"finished"}.
\item
  \texttt{\$n\_finished\_tasks} - Active binding that returns the
  current global count of finished tasks across all workers, commonly
  used as a termination criterion.
\end{itemize}

Both \texttt{\$push\_running\_tasks()} and \texttt{\$finish\_tasks()}
accept an \texttt{extra} argument for associating auxiliary information
with tasks and results that does not naturally fit into the inputs or
outputs.

\subsection{Tasks}\label{sec-tasks}

Tasks are the basic units through which workers exchange information.
Each task consists of four components: a unique key used to reference
the task in the \texttt{Redis} database, a computational state, an input
(\texttt{xs}), and an output (\texttt{ys}). The input and output are
lists that may contain arbitrary data.

Tasks always have an associated computational state: \texttt{"running"},
\texttt{"finished"}, \texttt{"failed"}, or \texttt{"queued"}. The
typical task lifecycle involves two transitions. The
\texttt{\$push\_running\_tasks()} method creates tasks marked as
\texttt{"running"} and returns their keys. Upon completion,
\texttt{\$finish\_tasks()} marks tasks as \texttt{"finished"} and stores
the associated results. Tasks that encounter errors can be marked as
\texttt{"failed"} using the \texttt{\$fail\_tasks()} method (see
Section~\ref{sec-error-handling}). The fourth state, \texttt{"queued"},
supports a queue mechanism described in Section~\ref{sec-queues}.

\subsection{Queues}\label{sec-queues}

In addition to the autonomous loop pattern, \texttt{rush} supports a
queue mechanism for cases in which tasks are created centrally and
distributed to workers. The \texttt{\$push\_tasks()} method creates
tasks in the \texttt{"queued"} state:

\begin{Shaded}
\begin{Highlighting}[]
\NormalTok{xss }\OtherTok{=} \FunctionTok{lapply}\NormalTok{(}\DecValTok{1}\SpecialCharTok{:}\DecValTok{20}\NormalTok{, }\ControlFlowTok{function}\NormalTok{(i) \{}
  \FunctionTok{list}\NormalTok{(}\AttributeTok{x1 =} \FunctionTok{runif}\NormalTok{(}\DecValTok{1}\NormalTok{, }\SpecialCharTok{{-}}\DecValTok{5}\NormalTok{, }\DecValTok{5}\NormalTok{), }\AttributeTok{x2 =} \FunctionTok{runif}\NormalTok{(}\DecValTok{1}\NormalTok{, }\SpecialCharTok{{-}}\DecValTok{5}\NormalTok{, }\DecValTok{5}\NormalTok{))}
\NormalTok{\})}

\NormalTok{rush}\SpecialCharTok{$}\FunctionTok{push\_tasks}\NormalTok{(}\AttributeTok{xss =}\NormalTok{ xss)}
\end{Highlighting}
\end{Shaded}

Workers consume queued tasks using \texttt{\$pop\_task()}, which
retrieves the next task, marks it as \texttt{"running"}, and returns it.
When the queue is empty, the method returns \texttt{NULL}, which serves
as the termination signal:

\begin{Shaded}
\begin{Highlighting}[]
\NormalTok{worker\_loop\_queue }\OtherTok{=} \ControlFlowTok{function}\NormalTok{(rush, fun) \{}
  \ControlFlowTok{repeat}\NormalTok{ \{}
\NormalTok{    task }\OtherTok{=}\NormalTok{ rush}\SpecialCharTok{$}\FunctionTok{pop\_task}\NormalTok{()}
    \ControlFlowTok{if}\NormalTok{ (}\FunctionTok{is.null}\NormalTok{(task)) }\ControlFlowTok{break}
\NormalTok{    ys }\OtherTok{=} \FunctionTok{list}\NormalTok{(}\AttributeTok{y =} \FunctionTok{fun}\NormalTok{(task}\SpecialCharTok{$}\NormalTok{xs}\SpecialCharTok{$}\NormalTok{x1, task}\SpecialCharTok{$}\NormalTok{xs}\SpecialCharTok{$}\NormalTok{x2))}
\NormalTok{    rush}\SpecialCharTok{$}\FunctionTok{finish\_tasks}\NormalTok{(task}\SpecialCharTok{$}\NormalTok{key, }\AttributeTok{yss =} \FunctionTok{list}\NormalTok{(ys))}
\NormalTok{  \}}
\NormalTok{\}}
\end{Highlighting}
\end{Shaded}

The two patterns can be combined: workers first drain a queue of
centrally created tasks, then transition to an autonomous loop for the
remainder of the computation.

\subsection{Error handling}\label{sec-error-handling}

\texttt{rush} provides error-handling mechanisms for two failure modes:
R errors during task evaluation and unexpected worker crashes. For
errors that occur during task evaluation, users must catch them within
the worker loop and mark the corresponding task as \texttt{"failed"}
using \texttt{\$fail\_tasks()}. If the error is not caught, the worker
will crash.

\begin{Shaded}
\begin{Highlighting}[]
\FunctionTok{tryCatch}\NormalTok{(\{}
\NormalTok{  ys }\OtherTok{=} \FunctionTok{list}\NormalTok{(}\AttributeTok{y =} \FunctionTok{fun}\NormalTok{(xs}\SpecialCharTok{$}\NormalTok{x1, xs}\SpecialCharTok{$}\NormalTok{x2))}
\NormalTok{  rush}\SpecialCharTok{$}\FunctionTok{finish\_tasks}\NormalTok{(key, }\AttributeTok{yss =} \FunctionTok{list}\NormalTok{(ys))}
\NormalTok{\}, }\AttributeTok{error =} \ControlFlowTok{function}\NormalTok{(e) \{}
\NormalTok{  condition }\OtherTok{=} \FunctionTok{list}\NormalTok{(}\AttributeTok{message =}\NormalTok{ e}\SpecialCharTok{$}\NormalTok{message)}
\NormalTok{  rush}\SpecialCharTok{$}\FunctionTok{fail\_tasks}\NormalTok{(key, }\AttributeTok{conditions =} \FunctionTok{list}\NormalTok{(condition))}
\NormalTok{\})}
\end{Highlighting}
\end{Shaded}

When a worker fails unexpectedly due to a crash, segmentation fault, or
lost network connection, its tasks may remain in the \texttt{"running"}
state indefinitely. The \texttt{\$detect\_lost\_workers()} method
identifies such workers and updates their state to
\texttt{"terminated"}. For workers started with
\texttt{\$start\_workers()} or \texttt{\$start\_local\_workers()}, lost
worker detection works automatically by checking process status. Workers
started via \texttt{\$worker\_script()} require an additional heartbeat
mechanism: a background process periodically refreshes a \texttt{Redis}
key with a set expiration timeout, and if the worker fails, the key
expires, signaling that the worker is lost:

\begin{Shaded}
\begin{Highlighting}[]
\NormalTok{script }\OtherTok{=}\NormalTok{ rush}\SpecialCharTok{$}\FunctionTok{worker\_script}\NormalTok{(}
  \AttributeTok{worker\_loop =}\NormalTok{ worker\_loop,}
  \AttributeTok{heartbeat\_period =} \DecValTok{1}\NormalTok{,}
  \AttributeTok{heartbeat\_expire =} \DecValTok{3}\NormalTok{)}
\FunctionTok{cat}\NormalTok{(}\FunctionTok{strwrap}\NormalTok{(script, }\AttributeTok{width =} \DecValTok{76}\NormalTok{, }\AttributeTok{exdent =} \DecValTok{2}\NormalTok{), }\AttributeTok{sep =} \StringTok{"}\SpecialCharTok{\textbackslash{}n}\StringTok{"}\NormalTok{)}
\end{Highlighting}
\end{Shaded}

\begin{verbatim}
Rscript -e 'rush::start_worker(network_id = "demo-network", config =
  list(scheme = "redis", host = "127.0.0.1", port = "6379"),
  heartbeat_period = 1L, heartbeat_expire = 3L)'
\end{verbatim}

\subsection{Logging}\label{logging}

Workers can write messages generated via the \texttt{lgr}
\citep{fleck2026} package to the database. The \texttt{lgr\_thresholds}
argument of \texttt{\$start\_workers()} specifies the logging level for
each logger by name, e.g., \texttt{c("mlr3/rush"\ =\ "debug")}. The
\texttt{mlr3/} prefix denotes the parent logger shared across the
\texttt{mlr3} ecosystem, into which \texttt{rush} integrates. Logging
introduces a minor performance overhead and is disabled by default.

\begin{Shaded}
\begin{Highlighting}[]
\NormalTok{wl\_log\_message }\OtherTok{=} \ControlFlowTok{function}\NormalTok{(rush) \{}
\NormalTok{  lg }\OtherTok{=}\NormalTok{ lgr}\SpecialCharTok{::}\FunctionTok{get\_logger}\NormalTok{(}\StringTok{"mlr3/rush"}\NormalTok{)}
\NormalTok{  lg}\SpecialCharTok{$}\FunctionTok{info}\NormalTok{(}\StringTok{"This is an info message from worker \%s"}\NormalTok{, rush}\SpecialCharTok{$}\NormalTok{worker\_id)}
\NormalTok{\}}

\NormalTok{rush}\SpecialCharTok{$}\FunctionTok{start\_workers}\NormalTok{(}
  \AttributeTok{worker\_loop =}\NormalTok{ wl\_log\_message,}
  \AttributeTok{n\_workers =} \DecValTok{2}\NormalTok{,}
  \AttributeTok{lgr\_thresholds =} \FunctionTok{c}\NormalTok{(}\StringTok{"mlr3/rush"} \OtherTok{=} \StringTok{"info"}\NormalTok{))}
\end{Highlighting}
\end{Shaded}

To retrieve all log entries, use the \texttt{\$read\_log()} method.

\begin{Shaded}
\begin{Highlighting}[]
\FunctionTok{Sys.sleep}\NormalTok{(}\DecValTok{1}\NormalTok{)}

\NormalTok{rush}\SpecialCharTok{$}\FunctionTok{read\_log}\NormalTok{()[, }\FunctionTok{list}\NormalTok{(worker\_id, msg)]}
\end{Highlighting}
\end{Shaded}

\begin{verbatim}
                         worker_id                             msg
                            <char>                          <char>
1: ungrammatical_beauceron_32fe... [rush] This is an info messa...
2:        accessible_bear_15e5052b [rush] This is an info messa...
\end{verbatim}

Outputs and messages can be written to files. Set a path to a folder via
the \texttt{message\_log} and \texttt{output\_log} arguments of
\texttt{\$start\_workers()}. Messages and outputs are written to files
named \texttt{message\_\textless{}worker\_id\textgreater{}.log} and
\texttt{output\_\textless{}worker\_id\textgreater{}.log} in that folder,
respectively.

\begin{Shaded}
\begin{Highlighting}[]
\NormalTok{rush}\SpecialCharTok{$}\FunctionTok{start\_workers}\NormalTok{(}
  \AttributeTok{worker\_loop =}\NormalTok{ wl\_log\_message,}
  \AttributeTok{n\_workers =} \DecValTok{2}\NormalTok{,}
  \AttributeTok{message\_log =} \StringTok{"logs"}\NormalTok{,}
  \AttributeTok{output\_log =} \StringTok{"logs"}\NormalTok{)}
\end{Highlighting}
\end{Shaded}

\subsection{Data storage}\label{sec-data-storage}

By default, \texttt{rush} stores all data in a \texttt{Redis} database.
\texttt{Redis} is an open-source, in-memory key-value store commonly
used as a database, cache, and message broker. It offers extremely low
latency and supports multiple data structures such as strings, hashes,
lists, and sets. \texttt{rush} maps its data model onto three
\texttt{Redis} data structures: hashes, sets, and lists.

\texttt{Redis} hashes are a data structure for storing associative
field--value pairs under a single key. They are optimized for memory
efficiency and fast access to structured data. This makes them suitable
for representing the input, output, and metadata of tasks. The key of
the hash identifies the task in \texttt{Redis} and \texttt{rush}. A task
is stored as a \texttt{Redis} hash with the following structure:

\begin{verbatim}
key : xs | ys | xs_extra
\end{verbatim}

The field--value pairs are written by different methods, e.g.,
\texttt{\$push\_running\_tasks()} writes \texttt{xs} and
\texttt{\$finish\_tasks()} writes \texttt{ys}. The values of the fields
are serialized lists or atomic values, e.g., deserializing the
\texttt{xs} field gives \texttt{list(x1\ =\ 1,\ x2\ =\ 2)}. This
structure allows a hash to be efficiently converted into a table row and
multiple hashes to be joined into a table. When retrieved, multiple
hashes are joined into a tabular format:

\begin{verbatim}
| key | x1 | x2 | y | timestamp |
| 1.. |  3 |  4 | 7 |  12:04:11 |
| 2.. |  1 |  4 | 5 |  12:04:12 |
| 3.. |  1 |  1 | 2 |  12:04:13 |
\end{verbatim}

\texttt{Redis} sets are unordered collections of unique elements that
support efficient membership tests. \texttt{rush} uses sets to track
task states; for example, the keys of all currently running tasks are
stored in the \texttt{"running\_tasks"} set.

Lists in \texttt{Redis} are ordered collections stored as linked lists
that support efficient insertion and removal at both ends. \texttt{rush}
uses lists for two purposes: managing the task queue and recording
finished tasks. For the queue, the list structure is a natural choice,
as it supports atomic push and pop operations. Storing the finished
tasks in a list gives them an order by time, enabling workers to cheaply
retrieve only the most recent results.

\texttt{rush} uses three mechanisms to coordinate concurrent access to
the database. First, \texttt{Redis} executes commands on a single
thread, making individual operations such as adding a key to a set or
appending a key to a list atomic. Second, operations involving several
keys are grouped into \texttt{MULTI}/\texttt{EXEC} transactions, which
\texttt{Redis} executes without interruption by other commands. For
example, \texttt{\$push\_running\_tasks()} adds new keys to the
\texttt{"running\_tasks"} set and the list of all tasks in a single
transaction. Third, potentially competing state transitions are
implemented as atomic \texttt{Lua} scripts. These include transitions by
\texttt{\$finish\_tasks()} and \texttt{\$fail\_tasks()}, insuring that a
task is not recorded in two states simultaneously.

To minimize the number of \texttt{Redis} round-trips when workers
repeatedly access many tasks (common in iterative algorithms), each
\texttt{Rush} instance keeps a local \texttt{data.table} cache of the
finished tasks it has already seen. On subsequent fetches, only the keys
after the current cache length are requested; the corresponding hashes
are read and appended to the cached table. This way, repeated queries
scale with the number of newly finished tasks instead of the total
result history.

The \texttt{large\_objects\_path} option allows large objects to be
stored on a shared file system accessible to all workers. This can
reduce network transfer overhead by allowing workers on the same node to
benefit from cached file reads.

\subsection{Retrieving results}\label{retrieving-results}

Once all computational tasks are completed, the results can be retrieved
from the database. As already explained, the
\texttt{\$fetch\_finished\_tasks()} method returns a \texttt{data.table}
containing the task key, input, and result.

\subsection{Testing}\label{sec-testing}

\texttt{rush} includes 105 unit tests written with \texttt{testthat}
\citep{wickham2011}. GitHub Actions runs \texttt{R\ CMD\ check},
including these tests, on Ubuntu with the release and development
versions of R. The tests cover the core functionality, error handling,
lost-worker detection, concurrent access, and caching behavior of the
package.

\section{Ask-and-tell and asynchronous decentralized Bayesian
optimization}\label{sec-case-study}

\subsection{The ask-and-tell pattern in
optimization}\label{sec-ask-tell}

The worker loop we discussed in Section~\ref{sec-worker-loop} directly
mirrors the \emph{ask-and-tell} interface pattern that has become the
dominant API design for black-box optimization frameworks. In this
pattern, the optimizer exposes two methods --- \texttt{\$ask()}, which
proposes the next candidate point(s), and \texttt{\$tell()}, which
receives the corresponding evaluation result(s) --- while the user
retains full control over the evaluation loop. In the worker loop
template, the trio of \texttt{\$fetch\_tasks\_with\_state()},
\texttt{compute\_task\_inputs()}, and \texttt{\$push\_running\_tasks()}
corresponds to the \emph{ask} step, while \texttt{\$finish\_tasks()}
corresponds to the \emph{tell} step. This design decouples the optimizer
from the objective function evaluation, allowing users to integrate
optimization into arbitrary workflows. Prominent implementations include
\texttt{Optuna}
\citep[\texttt{study.ask()}/\texttt{study.tell()},][]{akiba2019},
\texttt{Ax}
\citep[\texttt{get\_next\_trial()}/\texttt{complete\_trial()},][]{olson2025},
\texttt{SMAC3} \citep{lindauer2022}, and \texttt{scikit-optimize}
\citep{head2018}.

The ask-and-tell interface enables two fundamentally different
parallelization strategies for optimization, which differ in where the
optimization logic runs and how workers coordinate.

\textbf{Synchronous batch parallelism.} In this scenario, a single
central process runs the optimizer. In each iteration, the optimizer
asks for a batch of \(q\) candidate points, which are then evaluated in
parallel across \(q\) workers using a standard parallel backend (e.g.,
\texttt{future}, \texttt{mirai}, or \texttt{parallel}). Once all \(q\)
evaluations complete, the results are told back to the optimizer, which
updates its model and proposes the next batch. No shared state beyond
the optimizer is needed, but depending on the complexity of the
\texttt{\$ask()} step, the central optimizer can become a bottleneck.
Furthermore, synchronous batches waste resources when evaluation times
are heterogeneous, as all workers must wait for the slowest evaluation
in each batch to complete.

\textbf{Asynchronous decentralized parallelism.} In this scenario, each
worker runs its own local optimization and ask-and-tell loop: it reads
the shared history of completed (and possibly running) evaluations, asks
its local optimizer for a proposal, evaluates it, and tells the result
back to a shared store. Workers operate fully independently --- they do
not wait for each other and do not coordinate through a central
controller. This eliminates both the controller bottleneck and the
synchronization barrier. However, it requires infrastructure for
shared-state coordination: workers must be able to read each other's
results and register their own tasks in a concurrent, conflict-free
manner.

\subsection{Asynchronous decentralized Bayesian
optimization}\label{sec-adbo}

Bayesian optimization (BO) is a sample-efficient strategy for optimizing
expensive black-box functions. It maintains a probabilistic surrogate
model (typically a Gaussian process or random forest) fitted to all
completed evaluations, and selects the next evaluation point by
optimizing an acquisition function that balances exploration and
exploitation. In the standard sequential setting, a single process
alternates between fitting the surrogate, proposing a point, and
evaluating it. In ADBO \citep{egele2023}, this loop is replicated across
multiple workers, each running independently. Every worker reads the
shared history of evaluations from the database, fits its own local
surrogate model, and proposes the next point autonomously --- without
waiting for other workers or coordinating through a central controller.
For design points currently being evaluated on other workers (i.e.,
without a known objective value yet), a fake objective is imputed ---
usually a value between the minimum and maximum observed objective
values in the archive. This penalizes the proposal of further design
points near those currently under evaluation. Results are written back
to the shared store as they complete, so all workers benefit from each
other's evaluations. To encourage diversity among proposals, each worker
uses a stochastic acquisition function with different
exploration--exploitation trade-offs (e.g., different \(\lambda\) values
in a confidence bound criterion). This architecture eliminates the
central bottleneck that arises in synchronous or centralized parallel
BO, where a single process must fit the surrogate and propose points for
all workers; see Section~\ref{sec-benchmarks-bo}.

We use the Branin function, defined over the domain \(x_1 \in [-5, 10]\)
and \(x_2 \in [0, 15]\), as a toy objective:

\begin{Shaded}
\begin{Highlighting}[]
\NormalTok{branin }\OtherTok{=} \ControlFlowTok{function}\NormalTok{(x1, x2) \{}
\NormalTok{  (x2 }\SpecialCharTok{{-}} \FloatTok{5.1} \SpecialCharTok{/}\NormalTok{ (}\DecValTok{4} \SpecialCharTok{*}\NormalTok{ pi}\SpecialCharTok{\^{}}\DecValTok{2}\NormalTok{) }\SpecialCharTok{*}\NormalTok{ x1}\SpecialCharTok{\^{}}\DecValTok{2} \SpecialCharTok{+} \DecValTok{5} \SpecialCharTok{/}\NormalTok{ pi }\SpecialCharTok{*}\NormalTok{ x1 }\SpecialCharTok{{-}} \DecValTok{6}\NormalTok{)}\SpecialCharTok{\^{}}\DecValTok{2} \SpecialCharTok{+}
    \DecValTok{10} \SpecialCharTok{*}\NormalTok{ (}\DecValTok{1} \SpecialCharTok{{-}} \DecValTok{1} \SpecialCharTok{/}\NormalTok{ (}\DecValTok{8} \SpecialCharTok{*}\NormalTok{ pi)) }\SpecialCharTok{*} \FunctionTok{cos}\NormalTok{(x1) }\SpecialCharTok{+} \DecValTok{10}
\NormalTok{\}}
\end{Highlighting}
\end{Shaded}

Next, we define an optimizer that proposes new points for evaluation. It
takes as input an \texttt{archive} of running and completed tasks and
returns a new candidate point \texttt{xs} for evaluation. The archive is
represented as a \texttt{data.table} object. If the \texttt{archive} is
empty, the optimizer proposes a random point. Otherwise, it imputes
missing objective values for running tasks using the mean objective
value of completed tasks. It then fits a random forest surrogate model
with the \texttt{ranger} \citep{wright2017} R package and proposes a new
point by minimizing a lower confidence bound acquisition function
\(\hat\mu(x) - \lambda\,\hat\sigma(x)\). The exploration parameter
\(\lambda\) is passed as an argument; each worker draws its own value
from \(\mathrm{Exp}(1)\) once at the start of the worker loop, following
\citet{egele2023}. Assigning each worker a different \(\lambda\)
realizes the diversification mechanism of ADBO: independent workers
explore different exploration--exploitation trade-offs without any
coordination. Fitting the surrogate model becomes increasingly expensive
as the number of tasks in the \texttt{archive} grows. A detailed
understanding of this optimizer is not necessary to follow the
subsequent examples. Two properties are relevant for what follows: it
uses previously evaluated points from other workers to inform new
proposals, and proposal generation incurs a non-negligible computational
cost.

\begin{Shaded}
\begin{Highlighting}[]
\NormalTok{optimizer }\OtherTok{=} \ControlFlowTok{function}\NormalTok{(archive, lambda) \{}
  \ControlFlowTok{if}\NormalTok{ (}\FunctionTok{nrow}\NormalTok{(archive) }\SpecialCharTok{==} \DecValTok{0}\NormalTok{) \{}
    \FunctionTok{return}\NormalTok{(}\FunctionTok{list}\NormalTok{(}\AttributeTok{x1 =} \FunctionTok{runif}\NormalTok{(}\DecValTok{1}\NormalTok{, }\SpecialCharTok{{-}}\DecValTok{5}\NormalTok{, }\DecValTok{10}\NormalTok{), }\AttributeTok{x2 =} \FunctionTok{runif}\NormalTok{(}\DecValTok{1}\NormalTok{, }\DecValTok{0}\NormalTok{, }\DecValTok{15}\NormalTok{)))}
\NormalTok{  \}}

\NormalTok{  mean\_y }\OtherTok{=} \FunctionTok{mean}\NormalTok{(archive}\SpecialCharTok{$}\NormalTok{y, }\AttributeTok{na.rm =} \ConstantTok{TRUE}\NormalTok{)}
\NormalTok{  archive[}\StringTok{"running"}\NormalTok{, y }\SpecialCharTok{:}\ErrorTok{=}\NormalTok{ mean\_y, on }\OtherTok{=} \StringTok{"state"}\NormalTok{]}

\NormalTok{  surrogate }\OtherTok{=}\NormalTok{ ranger}\SpecialCharTok{::}\FunctionTok{ranger}\NormalTok{(}
\NormalTok{    y }\SpecialCharTok{\textasciitilde{}}\NormalTok{ x1 }\SpecialCharTok{+}\NormalTok{ x2,}
    \AttributeTok{data =}\NormalTok{ archive,}
    \AttributeTok{num.trees =} \DecValTok{100}\NormalTok{L,}
    \AttributeTok{keep.inbag =} \ConstantTok{TRUE}\NormalTok{)}

\NormalTok{  xdt }\OtherTok{=}\NormalTok{ data.table}\SpecialCharTok{::}\FunctionTok{data.table}\NormalTok{(}
    \AttributeTok{x1 =} \FunctionTok{runif}\NormalTok{(}\DecValTok{1000}\NormalTok{, }\SpecialCharTok{{-}}\DecValTok{5}\NormalTok{, }\DecValTok{10}\NormalTok{),}
    \AttributeTok{x2 =} \FunctionTok{runif}\NormalTok{(}\DecValTok{1000}\NormalTok{, }\DecValTok{0}\NormalTok{, }\DecValTok{15}\NormalTok{))}
\NormalTok{  p }\OtherTok{=} \FunctionTok{predict}\NormalTok{(surrogate, xdt, }\AttributeTok{type =} \StringTok{"se"}\NormalTok{, }\AttributeTok{se.method =} \StringTok{"jack"}\NormalTok{)}
\NormalTok{  cb }\OtherTok{=}\NormalTok{ p}\SpecialCharTok{$}\NormalTok{predictions }\SpecialCharTok{{-}}\NormalTok{ lambda }\SpecialCharTok{*}\NormalTok{ p}\SpecialCharTok{$}\NormalTok{se}

  \FunctionTok{as.list}\NormalTok{(xdt[}\FunctionTok{which.min}\NormalTok{(cb)])}
\NormalTok{\}}
\end{Highlighting}
\end{Shaded}

The BO worker loop combines the optimizer with the \texttt{rush} API.
Note that it uses \texttt{\$fetch\_tasks\_with\_state()} to retrieve
both running and finished tasks, allowing the optimizer to impute values
for evaluations still in progress. This is the ADBO pattern: each worker
fits its own surrogate model to all available data and independently
proposes the next configuration.

\begin{Shaded}
\begin{Highlighting}[]
\NormalTok{workerloop\_adbo }\OtherTok{=} \ControlFlowTok{function}\NormalTok{(rush, branin, optimizer) \{}
  \CommentTok{\# evaluate the initial design}
  \ControlFlowTok{repeat}\NormalTok{ \{}
\NormalTok{    task }\OtherTok{=}\NormalTok{ rush}\SpecialCharTok{$}\FunctionTok{pop\_task}\NormalTok{()}
    \ControlFlowTok{if}\NormalTok{ (}\FunctionTok{is.null}\NormalTok{(task)) }\ControlFlowTok{break}
\NormalTok{    ys }\OtherTok{=} \FunctionTok{list}\NormalTok{(}\AttributeTok{y =} \FunctionTok{branin}\NormalTok{(task}\SpecialCharTok{$}\NormalTok{xs}\SpecialCharTok{$}\NormalTok{x1, task}\SpecialCharTok{$}\NormalTok{xs}\SpecialCharTok{$}\NormalTok{x2))}
\NormalTok{    rush}\SpecialCharTok{$}\FunctionTok{finish\_tasks}\NormalTok{(task}\SpecialCharTok{$}\NormalTok{key, }\AttributeTok{yss =} \FunctionTok{list}\NormalTok{(ys))}
\NormalTok{  \}}

  \CommentTok{\# draw the worker{-}specific exploration parameter}
\NormalTok{  lambda }\OtherTok{=} \FunctionTok{rexp}\NormalTok{(}\DecValTok{1}\NormalTok{)}

  \CommentTok{\# ADBO loop}
  \ControlFlowTok{while}\NormalTok{ (rush}\SpecialCharTok{$}\NormalTok{n\_finished\_tasks }\SpecialCharTok{\textless{}} \DecValTok{100}\NormalTok{) \{}
\NormalTok{    archive }\OtherTok{=}\NormalTok{ rush}\SpecialCharTok{$}\FunctionTok{fetch\_tasks\_with\_state}\NormalTok{(}\AttributeTok{states =} \FunctionTok{c}\NormalTok{(}\StringTok{"running"}\NormalTok{, }\StringTok{"finished"}\NormalTok{))}
\NormalTok{    xs }\OtherTok{=} \FunctionTok{optimizer}\NormalTok{(archive, lambda)}

\NormalTok{    key }\OtherTok{=}\NormalTok{ rush}\SpecialCharTok{$}\FunctionTok{push\_running\_tasks}\NormalTok{(}\AttributeTok{xss =} \FunctionTok{list}\NormalTok{(xs))}

\NormalTok{    ys }\OtherTok{=} \FunctionTok{list}\NormalTok{(}\AttributeTok{y =} \FunctionTok{branin}\NormalTok{(xs}\SpecialCharTok{$}\NormalTok{x1, xs}\SpecialCharTok{$}\NormalTok{x2))}
\NormalTok{    rush}\SpecialCharTok{$}\FunctionTok{finish\_tasks}\NormalTok{(key, }\AttributeTok{yss =} \FunctionTok{list}\NormalTok{(ys))}
\NormalTok{  \}}
\NormalTok{\}}
\end{Highlighting}
\end{Shaded}

In practice, the optimization often starts with an initial design --- a
set of space-filling points that seed the surrogate model. The queue
mechanism from Section~\ref{sec-queues} provides a natural way to
distribute such an initial design: a structured design (e.g., Latin
hypercube sampling) is created once with the \texttt{lhs}
\citep{carnell2026} R package and pushed to the queue, and workers drain
it before transitioning to the model-based loop.

\begin{Shaded}
\begin{Highlighting}[]
\NormalTok{lhs\_points }\OtherTok{=}\NormalTok{ lhs}\SpecialCharTok{::}\FunctionTok{maximinLHS}\NormalTok{(}\AttributeTok{n =} \DecValTok{25}\NormalTok{, }\AttributeTok{k =} \DecValTok{2}\NormalTok{)}
\NormalTok{x1\_lower }\OtherTok{=} \SpecialCharTok{{-}}\DecValTok{5}
\NormalTok{x1\_range }\OtherTok{=} \DecValTok{15}
\NormalTok{x2\_lower }\OtherTok{=} \DecValTok{0}
\NormalTok{x2\_range }\OtherTok{=} \DecValTok{15}

\NormalTok{xss }\OtherTok{=} \FunctionTok{lapply}\NormalTok{(}\DecValTok{1}\SpecialCharTok{:}\DecValTok{25}\NormalTok{, }\ControlFlowTok{function}\NormalTok{(i) \{}
  \CommentTok{\# rescale to the domain}
  \FunctionTok{list}\NormalTok{(}
    \AttributeTok{x1 =}\NormalTok{ lhs\_points[i, }\DecValTok{1}\NormalTok{] }\SpecialCharTok{*}\NormalTok{ x1\_range }\SpecialCharTok{+}\NormalTok{ x1\_lower,}
    \AttributeTok{x2 =}\NormalTok{ lhs\_points[i, }\DecValTok{2}\NormalTok{] }\SpecialCharTok{*}\NormalTok{ x2\_range }\SpecialCharTok{+}\NormalTok{ x2\_lower)}
\NormalTok{\})}

\NormalTok{rush}\SpecialCharTok{$}\FunctionTok{push\_tasks}\NormalTok{(}\AttributeTok{xss =}\NormalTok{ xss)}
\end{Highlighting}
\end{Shaded}

\section{Benchmarking of core
functions}\label{benchmarking-of-core-functions}

\subsection{Experimental setup}\label{experimental-setup}

The evaluation of short-running tasks is sensitive to the overhead
introduced by \texttt{rush}. To quantify this overhead, we benchmark the
core functions: \texttt{\$push\_running\_tasks()},
\texttt{\$finish\_tasks()}, and \texttt{\$fetch\_finished\_tasks()}. For
\texttt{\$push\_running\_tasks()} and \texttt{\$finish\_tasks()}, we
vary the number of parameters (1, 10, 100) of the task and the payload
size (1, 10, 100, 1,000, 10,000) of each parameter. In the case of
\texttt{\$finish\_tasks()}, the payload size refers to the result. The
payload size is the number of elements in a double vector where each
double takes 8 bytes, i.e., a payload size of 10,000 corresponds to 80
kB. We benchmark \texttt{\$fetch\_finished\_tasks()} both with and
without caching to measure the effect of the caching mechanism. In
addition to the number of parameters and payload size, we vary the
number of tasks (1, 10, 100, 1,000, 10,000, 100,000). When caching is
enabled, the cache holds all previously fetched tasks except the most
recent one, so that each call retrieves only a single new result from
the database. For both \texttt{\$fetch\_finished\_tasks()} benchmarks,
we cap the payload size at 100 to avoid creating overly large tables in
memory. All code is available on zenodo at
https://zenodo.org/records/22899537 \citep{becker2026f}. All benchmarks
use \texttt{rush} version 1.3.0 with R 4.6.0 and \texttt{Redis} 8.10.1.
All measurements are taken by a single worker against a local
\texttt{Redis} instance communicating via a Unix socket.\footnote{The
  multi-node BO benchmarks in Section~\ref{sec-benchmarks-bo} use TCP,
  which adds some overhead.} The runtime is measured with the
\texttt{microbenchmark} \citep{mersmann2024} R package. We report the
median and the 95th percentile over the replicates. All figures are
created with the \texttt{ggplot2} \citep{wickham2016}, \texttt{cowplot}
\citep{wilke2025}, and \texttt{scales} \citep{wickham2025} R packages.

\subsection{Results}\label{results}

\begin{longtable}[]{@{}
  >{\raggedleft\arraybackslash}p{(\linewidth - 6\tabcolsep) * \real{0.2500}}
  >{\raggedleft\arraybackslash}p{(\linewidth - 6\tabcolsep) * \real{0.1912}}
  >{\raggedleft\arraybackslash}p{(\linewidth - 6\tabcolsep) * \real{0.2647}}
  >{\raggedleft\arraybackslash}p{(\linewidth - 6\tabcolsep) * \real{0.2941}}@{}}
\caption{Runtime of \protect\texttt{\$push\_running\_tasks()} and
\protect\texttt{\$finish\_tasks()} as a function of the number of fields
and payload size. For \protect\texttt{\$push\_running\_tasks()}, the
number of fields corresponds to the number of parameters and for
\protect\texttt{\$finish\_tasks()}, it corresponds to the number of
results. The runtime is reported as the median over 10,000 replicates
with the 95th percentile in
parentheses.}\label{tbl-runtime-push-finish-tasks}\tabularnewline
\toprule\noalign{}
\begin{minipage}[b]{\linewidth}\raggedleft
Number of Fields
\end{minipage} & \begin{minipage}[b]{\linewidth}\raggedleft
Payload Size
\end{minipage} & \begin{minipage}[b]{\linewidth}\raggedleft
Runtime Push (ms)
\end{minipage} & \begin{minipage}[b]{\linewidth}\raggedleft
Runtime Finish (ms)
\end{minipage} \\
\midrule\noalign{}
\endfirsthead
\toprule\noalign{}
\begin{minipage}[b]{\linewidth}\raggedleft
Number of Fields
\end{minipage} & \begin{minipage}[b]{\linewidth}\raggedleft
Payload Size
\end{minipage} & \begin{minipage}[b]{\linewidth}\raggedleft
Runtime Push (ms)
\end{minipage} & \begin{minipage}[b]{\linewidth}\raggedleft
Runtime Finish (ms)
\end{minipage} \\
\midrule\noalign{}
\endhead
\bottomrule\noalign{}
\endlastfoot
1 & 1 & 0.34 (0.75) & 0.26 (0.39) \\
1 & 10 & 0.33 (0.62) & 0.25 (0.34) \\
1 & 100 & 0.33 (0.73) & 0.26 (0.42) \\
1 & 1,000 & 0.34 (0.61) & 0.26 (0.32) \\
1 & 10,000 & 0.55 (1.27) & 0.34 (0.49) \\
10 & 1 & 0.33 (0.39) & 0.25 (0.32) \\
10 & 10 & 0.41 (1.15) & 0.25 (0.33) \\
10 & 100 & 0.44 (1.15) & 0.29 (0.38) \\
10 & 1,000 & 0.39 (1.16) & 0.35 (0.51) \\
10 & 10,000 & 1.89 (2.58) & 1.33 (2.03) \\
100 & 1 & 0.45 (1.11) & 0.27 (0.46) \\
100 & 10 & 0.53 (1.32) & 0.28 (0.36) \\
100 & 100 & 0.58 (1.30) & 0.36 (0.48) \\
100 & 1,000 & 0.92 (1.45) & 1.37 (2.07) \\
100 & 10,000 & 12.98 (25.41) & 12.07 (20.80) \\
\end{longtable}

The runtime of the functions \texttt{\$push\_running\_tasks()} and
\texttt{\$finish\_tasks()} is summarized in
Table~\ref{tbl-runtime-push-finish-tasks}. For payload sizes up to 100,
pushing a task to the database takes a median of 0.33--0.58 ms, and
finishing a task takes slightly less, 0.25--0.36 ms. For the largest
configuration (payload size 10,000 and 100 fields), the median push and
finish times increase to 12.98 ms and 12.07 ms, respectively. The
corresponding 95th percentiles are 25.41 ms and 20.80 ms, approximately
2.0 and 1.7 times the medians, indicating greater upper-tail latency for
large payloads. For realistic HPO workloads, which typically involve
tens of hyperparameters with scalar values (payload size 1), the median
per-task overhead remains well below 1 ms.

\begin{figure}

\centering{

\includegraphics[width=1\linewidth,height=\textheight,keepaspectratio,alt={Line plot of the runtime of fetching finished tasks with and without caching as a function of the total number of tasks.}]{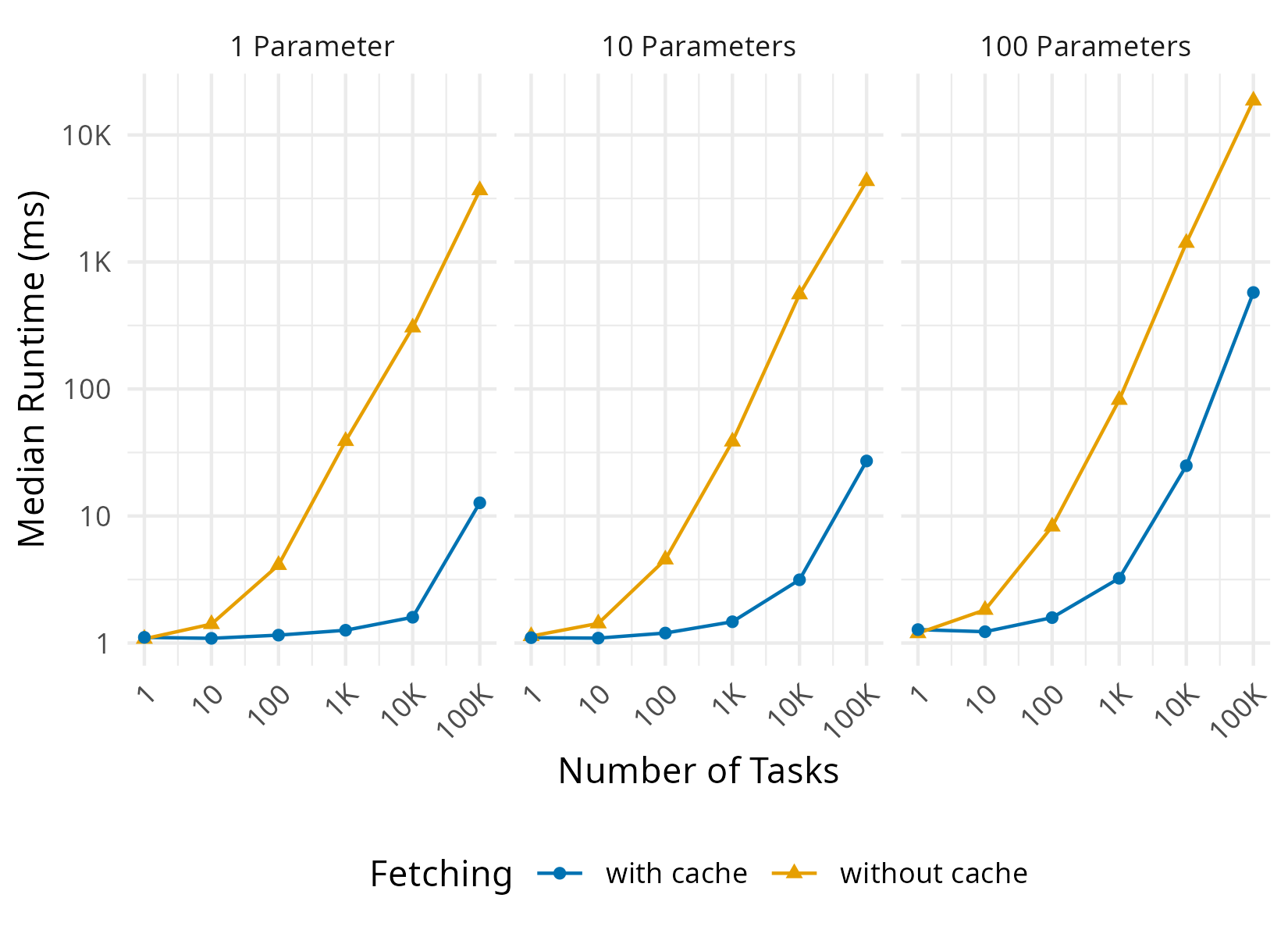}

}

\caption{\label{fig-fetch-finished-tasks}Median runtime in milliseconds
of fetching finished tasks with (blue circles) and without (orange
triangles) caching for different numbers of parameters as a function of
the total number of tasks in the database. Both axes are log-scaled. The
payload size is 1 in this benchmark. Without caching, all tasks are
retrieved from the database on each call. With caching, only one new
task is retrieved from the database; all previously fetched tasks are
served from the worker's cache.}

\end{figure}

For fetching, \texttt{rush} provides a built-in caching mechanism. The
difference between fetching with and without caching is negligible for a
small number of tasks. However, as the archive grows beyond 1,000 tasks,
caching yields a substantial speedup, as shown in
Figure~\ref{fig-fetch-finished-tasks} and
Table~\ref{tbl-runtime-finished-tasks}. At 10,000 tasks and 10
parameters, fetching without the cache takes approximately 556 ms,
compared to 3 ms with caching.

\section{Benchmarks of Bayesian optimization}\label{sec-benchmarks-bo}

\subsection{Experimental
setup}\label{sec-benchmarks-bo-experimental-setup}

We compare the effective CPU utilization of three parallelization
strategies for BO on HPO tasks, representing increasing degrees of
decentralization. We optimize nine hyperparameters of \texttt{LightGBM}
\citep{ke2017}, using the implementation in \texttt{mlr3extralearners}
\citep{fischer2025}, on four datasets, as summarized in
Table~\ref{tbl-lightgbm-hyperparameters} and Table~\ref{tbl-datasets}.
The datasets are downloaded from OpenML with \texttt{mlr3oml}
\citep{lang2026} and differ in their number of features and observations
to induce a range of training runtimes. Each experiment runs for 10
minutes on 448 workers. The benchmarks are conducted on a Linux cluster
at the Leibniz Supercomputing Centre (LRZ). Each node has 112 Intel Xeon
Sapphire Rapids cores and 488 GiB of memory, running SUSE Linux
Enterprise Server 15 SP6. We use four nodes for the benchmarks,
resulting in 448 cores in total. All code is available on zenodo at
https://zenodo.org/records/22899537 \citep{becker2026f}. All benchmarks
use \texttt{rush} version 1.3.0.

\textbf{Synchronous batch BO (CL).} Evaluates \(q\) hyperparameter
configurations in parallel using the constant liar (CL) approach
\citep{ginsbourger2010}. The first point is obtained by maximizing the
acquisition function. A fake value is then imputed, the surrogate model
is updated, and the next point is proposed by maximizing the acquisition
function again. Common imputation choices are the minimum, maximum, or
predicted posterior mean of already evaluated points. In our setup, CL
proposes 44 configurations evaluated with 10-fold cross-validation,
resulting in 440 parallel evaluations. The remaining 8 cores cannot be
used by design, since each configuration occupies 10 cores.
\texttt{mlr3} uses \texttt{mirai} to parallelize over the resampling
iterations. Batch parallelization requires all workers to finish before
the next batch can be proposed. We use early stopping to determine the
optimal number of boosting iterations, which naturally induces
heterogeneous runtimes: some models train for only a few iterations
while others run for up to 5,000. This heterogeneity exposes the
synchronization overhead inherent in batch approaches, as fast workers
remain idle while waiting for slower ones to complete.

\textbf{Asynchronous centralized BO (ACBO).} Eliminates synchronization
overhead by evaluating configurations asynchronously. When a worker
becomes idle, the central controller proposes a new configuration and
sends it to the worker without waiting for other evaluations to
complete. Objective values for configurations currently under evaluation
are imputed, as in CL. However, surrogate fitting and acquisition
function optimization remain centralized and are executed sequentially
by the controller. With many workers, this sequential bottleneck limits
scalability: new configurations cannot be proposed fast enough to keep
all workers occupied.

\textbf{Asynchronous decentralized BO (ADBO).} Distributes both
surrogate updates and acquisition function optimization across the
workers. Each worker maintains its own surrogate model and independently
proposes the next configuration. When a worker completes an evaluation,
it asynchronously shares the result via the database; other workers
incorporate this information into their local models. As in CL and ACBO,
objective values for configurations currently being evaluated are
imputed. ADBO additionally uses stochastic acquisition functions to
promote varying exploration--exploitation trade-offs between workers,
e.g., by initializing the \(\lambda\) parameter of the confidence bound
acquisition function with different values for each worker.\footnote{The
  workers therefore do not repeat each other's computations: their
  archives usually differ in the configurations under evaluation and the
  imputed values, and even on identical archives, the randomness in the
  surrogate, the acquisition function, and its optimizer leads to
  different proposals.} ADBO requires infrastructure that allows workers
to exchange information asynchronously without a central controller ---
the architecture provided by \texttt{rush}.

All three strategies are implemented with \texttt{mlr3tuning}
\citep{becker2025a} and \texttt{mlr3mbo} \citep{becker2026c} and use the
\texttt{mlr3mbo} defaults: a random forest surrogate model
\citep{wright2017}, a confidence bound acquisition function, and local
search for acquisition function optimization. The strategies differ in
the number of cores used per configuration. CL distributes the 10
resampling iterations across 10 cores and evaluates 44 configurations in
parallel. ACBO and ADBO execute the resampling iterations sequentially
on a single core, allowing up to 448 configurations to be evaluated
concurrently. For each task and replication, we randomly sample an
initial design of 100 configurations from the search space in
Table~\ref{tbl-lightgbm-hyperparameters} and evaluate them using 10-fold
cross-validation. The evaluation counts, wall-clock times, and
utilization in Table~\ref{tbl-benchmarks} and the traces in
Figure~\ref{fig-performance-algorithms} exclude the initial design. Each
task--algorithm combination is run five times. Benchmark jobs use fixed
random seeds, with initial designs shared across strategies and fixed
OpenML cross-validation splits. In asynchronous execution, result
ordering depends on timing, so fixed seeds alone do not ensure exact
reproducibility.

We measure effective CPU utilization following \citet{egele2023}. Let
\(n_\text{workers}\) denote the number of workers, where each worker
corresponds to a single CPU core, and \(T_\text{walltime}\) be the
elapsed wall-clock time. The total available CPU time is
\(T_\text{CPU} = T_\text{walltime} \cdot n_\text{workers}\). Let
\(T_\text{optimization}\) denote the cumulative time spent on surrogate
fitting, acquisition optimization, and model training, summed across all
workers. The effective CPU utilization is then
\(U = \frac{T_\text{optimization}}{T_\text{CPU}}\). Utilization thus
counts the time spent proposing configurations as well as evaluating
them. All three strategies fit one surrogate and optimize the
acquisition function once per proposed configuration. To separate the
two components, Table~\ref{tbl-benchmarks-extra} additionally reports
the learner utilization, which counts only the time spent evaluating
configurations.

\begin{longtable}[]{@{}
  >{\raggedright\arraybackslash}p{(\linewidth - 8\tabcolsep) * \real{0.2436}}
  >{\raggedright\arraybackslash}p{(\linewidth - 8\tabcolsep) * \real{0.1282}}
  >{\raggedleft\arraybackslash}p{(\linewidth - 8\tabcolsep) * \real{0.2564}}
  >{\raggedleft\arraybackslash}p{(\linewidth - 8\tabcolsep) * \real{0.1667}}
  >{\raggedleft\arraybackslash}p{(\linewidth - 8\tabcolsep) * \real{0.2051}}@{}}
\caption{Effective CPU utilization and the number of completed
evaluations for CL, ACBO, and ADBO across four benchmark tasks. For each
task--algorithm combination, the table reports the mean runtime of the
10-fold cross-validation (CV) in seconds, the number of completed
evaluations, and the resulting effective CPU utilization in percent. All
entries are the mean over five replications with the standard deviation
in parentheses. Because each algorithm follows a different search
trajectory, the configurations evaluated differ, leading to variation in
mean CV runtimes across algorithms on the same dataset. This complicates
direct comparison of raw evaluation counts: e.g., on airlines, ACBO's
mean evaluation (158 s) is shorter than CL's (286 s), so part of ACBO's
higher evaluation count reflects faster individual evaluations rather
than only better scheduling. The same effect explains the large standard
deviation of the evaluation count of ADBO on airlines, whose mean
evaluation ranges from 171 s to 403 s across
replications.}\label{tbl-benchmarks}\tabularnewline
\toprule\noalign{}
\begin{minipage}[b]{\linewidth}\raggedright
Task
\end{minipage} & \begin{minipage}[b]{\linewidth}\raggedright
Algorithm
\end{minipage} & \begin{minipage}[b]{\linewidth}\raggedleft
Mean Runtime CV (s)
\end{minipage} & \begin{minipage}[b]{\linewidth}\raggedleft
Evaluations
\end{minipage} & \begin{minipage}[b]{\linewidth}\raggedleft
Utilization (\%)
\end{minipage} \\
\midrule\noalign{}
\endfirsthead
\toprule\noalign{}
\begin{minipage}[b]{\linewidth}\raggedright
Task
\end{minipage} & \begin{minipage}[b]{\linewidth}\raggedright
Algorithm
\end{minipage} & \begin{minipage}[b]{\linewidth}\raggedleft
Mean Runtime CV (s)
\end{minipage} & \begin{minipage}[b]{\linewidth}\raggedleft
Evaluations
\end{minipage} & \begin{minipage}[b]{\linewidth}\raggedleft
Utilization (\%)
\end{minipage} \\
\midrule\noalign{}
\endhead
\bottomrule\noalign{}
\endlastfoot
credit-g & CL & 0.65 (0.06) & 502 (24) & 0.3 (0.0) \\
& ACBO & 0.53 (0.02) & 545 (13) & 0.3 (0.0) \\
& ADBO & 0.71 (0.02) & 15,796 (231) & 94.6 (0.1) \\
KDDCup09\_appetency & CL & 26.46 (2.71) & 132 (0) & 1.1 (0.1) \\
& ACBO & 23.21 (0.94) & 526 (7) & 4.9 (0.2) \\
& ADBO & 30.05 (0.80) & 6,520 (106) & 93.7 (0.1) \\
adult & CL & 11.53 (0.74) & 370 (24) & 1.8 (0.1) \\
& ACBO & 11.43 (2.69) & 500 (15) & 2.4 (0.6) \\
& ADBO & 15.12 (0.33) & 9,728 (108) & 95.7 (0.0) \\
airlines & CL & 285.90 (73.31) & 97 (20) & 9.2 (1.6) \\
& ACBO & 157.83 (29.29) & 334 (31) & 31.9 (7.6) \\
& ADBO & 270.85 (95.37) & 494 (365) & 99.5 (0.2) \\
\end{longtable}

\subsection{Results}\label{results-1}

The effective CPU utilization and the number of completed evaluations
are summarized in Table~\ref{tbl-benchmarks} as means and standard
deviations over the five replications. Across all tasks and all
replications, ADBO achieves substantially higher utilization than both
CL and ACBO, and this improvement is consistent across datasets with
both short and long training runtimes. The utilization of ADBO varies by
less than one percentage point between replications of the same task, so
the standard deviations in Table~\ref{tbl-benchmarks} are small compared
to the differences between the algorithms. ADBO also completes
substantially more evaluations within the same wall-clock budget.
Utilization includes the surrogate fitting needed to propose
configurations: on credit-g, ADBO spends a mean cumulative 211,614 s
fitting surrogates versus 11,164 s evaluating configurations
(Table~\ref{tbl-benchmarks-extra}). ADBO spends more time per proposal
on surrogate fitting than CL and ACBO because the cost of fitting a
random forest grows with the size of the archive, which reaches over
15,000 evaluations for ADBO but stays below 650 for CL and ACBO.
Counting only the time spent evaluating configurations, ADBO reaches a
mean learner utilization of 4.2\% on credit-g, compared to 0.1\% for CL
and ACBO, and between 56.2\% and 99.2\% on the other tasks, compared to
at most 31.7\% for CL and ACBO (Table~\ref{tbl-benchmarks-extra}).

All three methods reach similar final classification errors; ADBO
achieves slightly lower mean errors, with differences of up to 0.8
percentage points on credit-g and less than 0.1 percentage points on the
other tasks. High utilization alone therefore does not imply better
optimization performance. The additional evaluations mainly help reach a
given error earlier (Figure~\ref{fig-performance-algorithms}).

\begin{figure}

\centering{

\includegraphics[width=1\linewidth,height=\textheight,keepaspectratio,alt={Bar plots of CPU utilization over time for short-running and long-running objective functions.}]{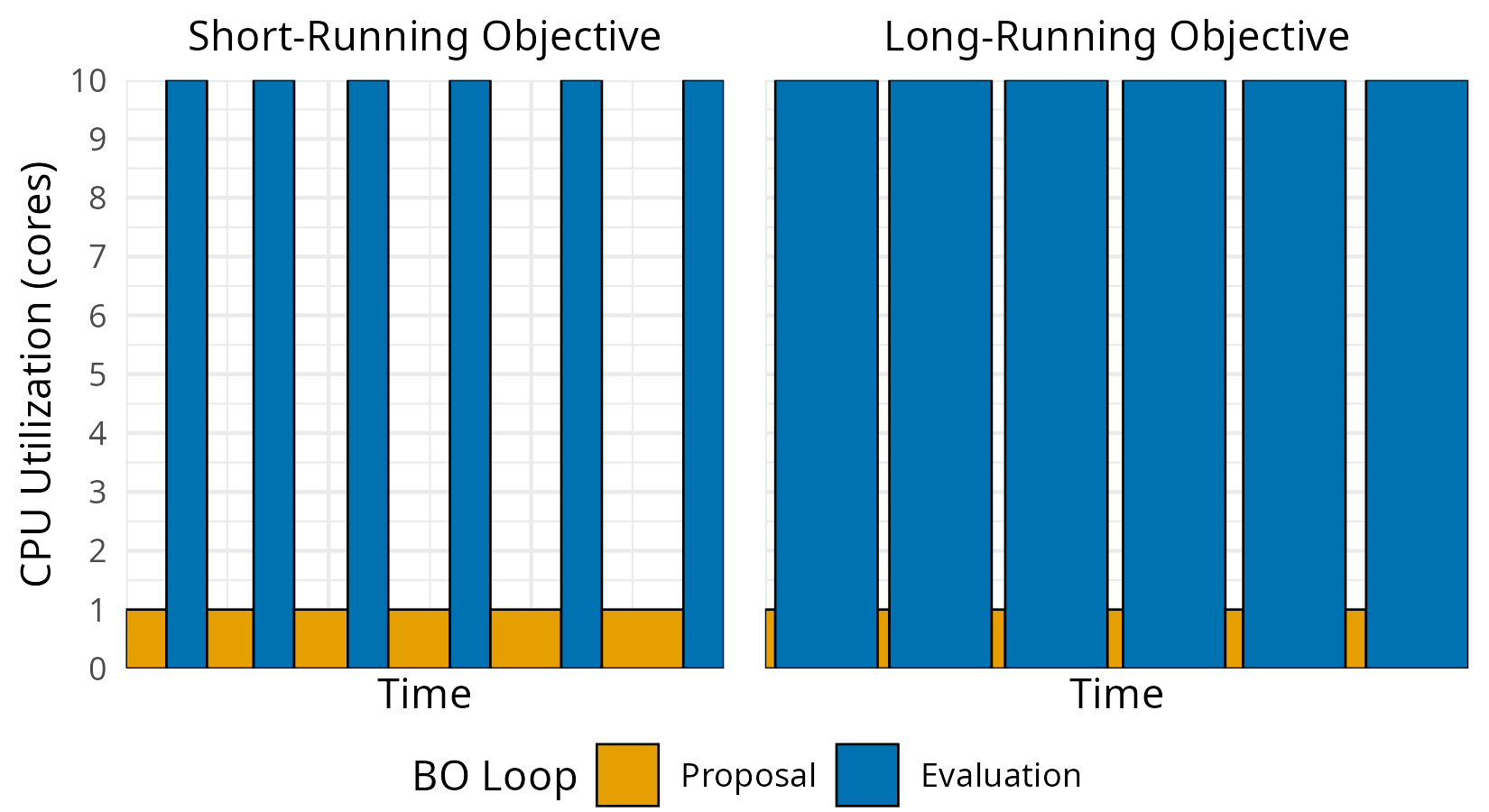}

}

\caption{\label{fig-cl-cpu-utilization}CPU utilization for short-running
and long-running objective functions. Bars represent the proposal phase
(light orange) and evaluation phase (dark blue) of the BO loop. Bar
height indicates CPU utilization on a 10-core machine: during proposal,
CL uses a single core, whereas during evaluation it uses all 10 cores.
The proposal phase grows over iterations as surrogate fitting becomes
more expensive with an increasing number of observations. For
long-running objectives, evaluation dominates wall-clock time, yielding
higher average CPU utilization than for short-running objectives. The
effect is far more pronounced at the 448-worker scale used in our
benchmarks, where the proposal phase occupies only a single worker,
about \(0.2\%\) of the available resources, leaving the remaining
\(99.8\%\) idle.}

\end{figure}

On tasks with very short training runtimes, centralized methods leave
most compute resources idle. This effect is most pronounced on the
credit-g task, where CL and ACBO complete only a few hundred evaluations
while ADBO completes over 15,000 in every replication. The utilization
of ADBO is between 270 and 300 times that of CL and ACBO across the five
replications. Figure~\ref{fig-cl-cpu-utilization} illustrates the
underlying mechanism. During the proposal phase, only a single core is
busy, and for short-running objectives this phase dominates the
wall-clock time.

On medium-cost tasks such as adult and KDDCup09\_appetency, the same
pattern persists. ACBO improves utilization relative to CL by
eliminating synchronization overhead, but the centralized surrogate
update remains a limiting factor: on adult, for instance, ACBO achieves
a mean utilization of 2.4\% compared to 95.7\% for ADBO.

The airlines task has the longest training times, which reduces the
relative overhead of surrogate updates and consequently increases
utilization for all algorithms. Utilization also varies most between
replications on this task, since a run evaluates only a few hundred
configurations and their cost differs strongly: the utilization of ACBO
ranges from 20\% to 41\% and its mean evaluation runtime from 132 s to
202 s across the five replications. Even in this favorable regime, ADBO
maintains a substantial advantage, achieving between 2.4 and 4.9 times
the utilization of ACBO in the individual replications. The
decentralized architecture therefore scales robustly across workload
regimes.

Beyond utilization, the parallelization strategy also affects budget
adherence. The wall-clock runtime of CL can overshoot the nominal
10-minute budget, sometimes substantially (e.g.~on KDDCup09\_appetency;
see Table~\ref{tbl-benchmarks-extra}), because CL checks the termination
criterion only after an entire batch finishes. If a batch starts shortly
before the time limit, the experiment continues until all evaluations
complete, introducing significant budget overruns. In contrast, ACBO and
ADBO check for termination after every completed evaluation, enforcing
the wall-clock constraint more accurately. Asynchronous scheduling
therefore improves both utilization and budget adherence.

\section{Discussion and conclusion}\label{discussion-and-conclusion}

We presented \texttt{rush}, a package for asynchronous and decentralized
optimization in R. Its database-centric architecture eliminates the
central controller bottleneck by allowing workers to communicate through
a shared \texttt{Redis} database and independently execute their own
optimization loops. The package combines sub-millisecond per-task
overhead with an efficient caching mechanism and integrates with the
\texttt{mlr3} ecosystem as the backend for asynchronous optimization in
\texttt{bbotk} and \texttt{mlr3tuning}. Our benchmarks demonstrate that
ADBO on \texttt{rush} achieves a CPU utilization above 93\% on 448
workers in every run, compared to single-digit utilization for
centralized approaches on short-running tasks. The sub-millisecond
overhead measurements apply to a single worker using a local Unix
socket; per-operation latency under concurrent TCP access was not
measured separately.

\begin{figure}

\centering{

\includegraphics[width=1\linewidth,height=\textheight,keepaspectratio,alt={Line plot of the incumbent classification error over wall-clock time for CL, ACBO, and ADBO on the four benchmark tasks.}]{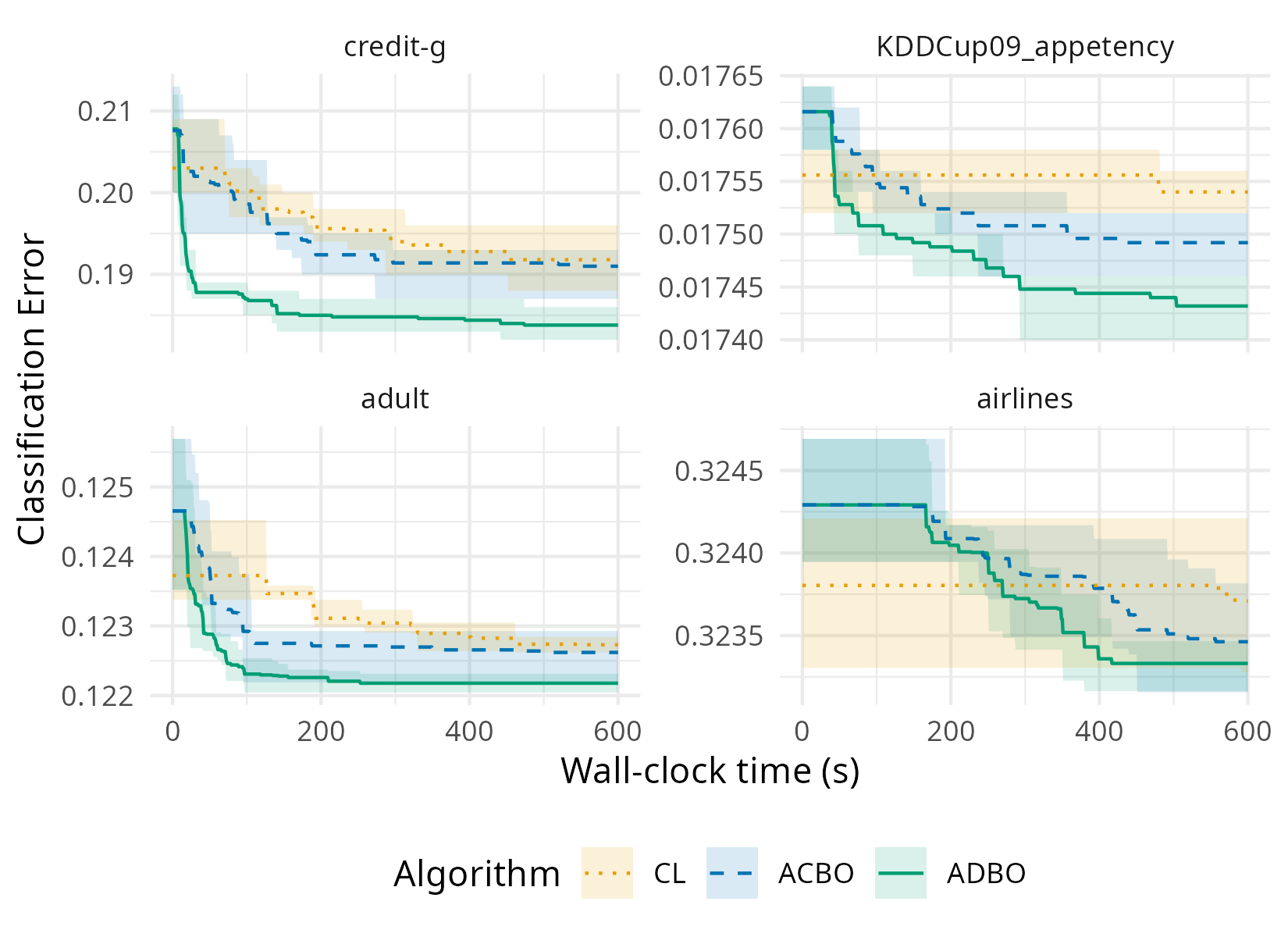}

}

\caption{\label{fig-performance-algorithms}Incumbent classification
error over wall-clock time of CL (dotted, orange), ACBO (dashed, blue),
and ADBO (solid, green). Lines show the mean incumbent over the five
replications. The shaded bands go from the best to the worst
replication. All algorithms start from the same initial design of 100
configurations, evaluated before the timed runs
(Section~\ref{sec-benchmarks-bo-experimental-setup}). Each trace begins
with the first completed evaluation proposed by the BO loop. For CL this
is the completion of the first batch of 44 configurations. ACBO and ADBO
update the incumbent after every completed evaluation.}

\end{figure}

Despite these results, several limitations should be noted. The
benchmarks are designed to measure utilization and not final
optimization performance. All three algorithms reach similar final
classification errors within the 10-minute budget
(Figure~\ref{fig-performance-algorithms} and
Table~\ref{tbl-benchmarks-extra}). ADBO reaches the lowest mean error on
all four tasks, but the differences are small: 0.8 percentage points on
credit-g and less than 0.1 percentage points on the other three tasks.
The additional evaluations under ADBO mainly translate into reaching a
given error earlier rather than into a lower final error. Whether the
higher throughput yields better final solutions on more complex problems
like hierarchical search spaces must be evaluated in future work. The
caching mechanism reduces fetch overhead substantially, but the cost of
binding new rows to the cached \texttt{data.table} grows with archive
size, which may become relevant for optimization runs with very large
numbers of evaluations. We do not provide a direct comparison with
Python frameworks such as \texttt{DeepHyper} or \texttt{Optuna}, which
would help position \texttt{rush} in the broader cross-language
landscape. Worker startup can be delayed by transferring large datasets
over the network, as observed for \texttt{kddcup09\_appetency} on the
LRZ cluster. Where a shared file system is available, the
\texttt{large\_objects\_path} option can reduce this overhead through
cached file reads (Section~\ref{sec-data-storage}).

Several directions for future work emerge from these limitations.
Adaptive caching strategies, such as pre-allocation, could mitigate the
growing cost of row binding for very large archives. The queue mechanism
could be extended to support more complex task-dependency graphs,
enabling workflows in which tasks have prerequisites or conditional
branching.

By providing a framework for decentralized optimization in R,
\texttt{rush} enables a class of algorithms that was previously
available only in the Python ecosystem. As computational budgets and
model complexity continue to grow, the ability to scale optimization
efficiently across large numbers of workers becomes increasingly
important. We hope that \texttt{rush} will serve as a foundation for the
development and application of new distributed optimization algorithms
in R.

\section*{Acknowledgments}\label{acknowledgments}
\addcontentsline{toc}{section}{Acknowledgments}

The authors gratefully acknowledge the computational and data resources
provided by the Leibniz Supercomputing Centre (www.lrz.de). We thank
Martin Binder and Sebastian Fischer for their valuable feedback.

\section*{Competing interests}\label{competing-interests}
\addcontentsline{toc}{section}{Competing interests}

The authors develop and maintain the \texttt{rush}, \texttt{bbotk},
\texttt{mlr3tuning}, and \texttt{mlr3mbo} packages benchmarked in this
paper. They declare no financial competing interests.

\section*{References}\label{references}
\addcontentsline{toc}{section}{References}

\renewcommand{\bibsection}{}
\bibliography{rush.bib}

\newpage

\section*{Appendix}\label{appendix}
\addcontentsline{toc}{section}{Appendix}

\begin{longtable}[]{@{}
  >{\raggedleft\arraybackslash}p{(\linewidth - 8\tabcolsep) * \real{0.1584}}
  >{\raggedleft\arraybackslash}p{(\linewidth - 8\tabcolsep) * \real{0.3069}}
  >{\raggedleft\arraybackslash}p{(\linewidth - 8\tabcolsep) * \real{0.1287}}
  >{\raggedleft\arraybackslash}p{(\linewidth - 8\tabcolsep) * \real{0.2178}}
  >{\raggedleft\arraybackslash}p{(\linewidth - 8\tabcolsep) * \real{0.1881}}@{}}
\caption{Runtime of \protect\texttt{\$fetch\_finished\_tasks()} with and
without caching as a function of the number of tasks, number of
parameters or results, and payload size. When caching is enabled, the
cache holds all previously fetched tasks except the most recent one, so
that each call retrieves only a single new result from the database. The
runtime is reported as the median with the 95th percentile in
parentheses over 1,000 replicates for up to 1,000 tasks, 100 replicates
for 10,000 tasks, and 10 replicates for 100,000
tasks.}\label{tbl-runtime-finished-tasks}\tabularnewline
\toprule\noalign{}
\begin{minipage}[b]{\linewidth}\raggedleft
Number of Tasks
\end{minipage} & \begin{minipage}[b]{\linewidth}\raggedleft
Number of Parameters / Results
\end{minipage} & \begin{minipage}[b]{\linewidth}\raggedleft
Payload Size
\end{minipage} & \begin{minipage}[b]{\linewidth}\raggedleft
Runtime (ms)
\end{minipage} & \begin{minipage}[b]{\linewidth}\raggedleft
Runtime Cache (ms)
\end{minipage} \\
\midrule\noalign{}
\endfirsthead
\toprule\noalign{}
\begin{minipage}[b]{\linewidth}\raggedleft
Number of Tasks
\end{minipage} & \begin{minipage}[b]{\linewidth}\raggedleft
Number of Parameters / Results
\end{minipage} & \begin{minipage}[b]{\linewidth}\raggedleft
Payload Size
\end{minipage} & \begin{minipage}[b]{\linewidth}\raggedleft
Runtime (ms)
\end{minipage} & \begin{minipage}[b]{\linewidth}\raggedleft
Runtime Cache (ms)
\end{minipage} \\
\midrule\noalign{}
\endhead
\bottomrule\noalign{}
\endlastfoot
1 & 1 & 1 & 1.08 (3.19) & 1.11 (3.26) \\
1 & 1 & 10 & 1.14 (3.27) & 1.07 (3.28) \\
1 & 1 & 100 & 1.13 (3.12) & 1.07 (3.00) \\
1 & 10 & 1 & 1.13 (3.32) & 1.10 (3.29) \\
1 & 10 & 10 & 1.09 (2.84) & 1.19 (3.05) \\
1 & 10 & 100 & 1.10 (3.37) & 1.15 (2.65) \\
1 & 100 & 1 & 1.19 (3.59) & 1.28 (3.18) \\
1 & 100 & 10 & 1.20 (3.52) & 1.23 (3.63) \\
1 & 100 & 100 & 1.39 (3.78) & 1.33 (3.86) \\
10 & 1 & 1 & 1.41 (3.74) & 1.09 (3.24) \\
10 & 1 & 10 & 1.43 (3.81) & 1.10 (3.33) \\
10 & 1 & 100 & 1.44 (3.49) & 1.12 (3.24) \\
10 & 10 & 1 & 1.43 (3.86) & 1.09 (3.30) \\
10 & 10 & 10 & 1.53 (2.52) & 1.10 (2.87) \\
10 & 10 & 100 & 1.52 (1.95) & 1.11 (3.39) \\
10 & 100 & 1 & 1.82 (3.31) & 1.23 (3.66) \\
10 & 100 & 10 & 1.98 (3.69) & 1.28 (1.50) \\
10 & 100 & 100 & 2.47 (7.73) & 1.30 (2.52) \\
100 & 1 & 1 & 4.11 (9.54) & 1.15 (2.49) \\
100 & 1 & 10 & 4.44 (11.13) & 1.15 (3.22) \\
100 & 1 & 100 & 4.40 (9.63) & 1.16 (3.08) \\
100 & 10 & 1 & 4.55 (9.44) & 1.20 (2.52) \\
100 & 10 & 10 & 4.90 (10.67) & 1.19 (3.19) \\
100 & 10 & 100 & 5.48 (11.28) & 1.23 (3.33) \\
100 & 100 & 1 & 8.26 (14.24) & 1.59 (4.17) \\
100 & 100 & 10 & 9.35 (17.08) & 1.50 (4.03) \\
100 & 100 & 100 & 17.47 (137.19) & 1.60 (4.20) \\
1,000 & 1 & 1 & 38.93 (57.40) & 1.26 (3.72) \\
1,000 & 1 & 10 & 37.59 (49.12) & 1.29 (3.22) \\
1,000 & 1 & 100 & 38.57 (57.29) & 1.29 (3.47) \\
1,000 & 10 & 1 & 38.57 (50.38) & 1.47 (3.21) \\
1,000 & 10 & 10 & 43.13 (57.46) & 1.43 (3.93) \\
1,000 & 10 & 100 & 49.16 (184.47) & 1.49 (3.88) \\
1,000 & 100 & 1 & 82.07 (91.55) & 3.24 (5.53) \\
1,000 & 100 & 10 & 100.34 (148.00) & 3.25 (6.13) \\
1,000 & 100 & 100 & 308.26 (360.19) & 2.78 (6.33) \\
10,000 & 1 & 1 & 305.06 (351.18) & 1.60 (4.23) \\
10,000 & 1 & 10 & 466.67 (524.38) & 1.67 (4.42) \\
10,000 & 1 & 100 & 331.99 (507.94) & 1.74 (4.53) \\
10,000 & 10 & 1 & 555.92 (581.49) & 3.15 (6.68) \\
10,000 & 10 & 10 & 436.40 (673.95) & 3.80 (9.14) \\
10,000 & 10 & 100 & 825.37 (1,015.53) & 8.42 (16.69) \\
10,000 & 100 & 1 & 1,413.02 (1,564.17) & 24.88 (486.43) \\
10,000 & 100 & 10 & 1,377.38 (1,870.65) & 25.06 (96.38) \\
10,000 & 100 & 100 & 2,204.29 (2,528.53) & 24.93 (748.25) \\
100,000 & 1 & 1 & 3,667.42 (3,997.82) & 12.70 (37.58) \\
100,000 & 1 & 10 & 3,662.08 (4,210.95) & 6.72 (27.55) \\
100,000 & 1 & 100 & 4,120.23 (4,480.65) & 20.10 (77.98) \\
100,000 & 10 & 1 & 4,331.54 (4,750.86) & 27.12 (62.05) \\
100,000 & 10 & 10 & 5,497.61 (5,826.04) & 40.10 (72.11) \\
100,000 & 10 & 100 & 6,611.68 (7,147.28) & 41.13 (64.47) \\
100,000 & 100 & 1 & 18,637.08 (22,791.05) & 575.03 (7,251.52) \\
100,000 & 100 & 10 & 22,156.63 (26,338.94) & 294.34 (2,407.24) \\
100,000 & 100 & 100 & 25,561.06 (29,777.61) & 335.69 (422.81) \\
\end{longtable}

\begin{longtable}[]{@{}llll@{}}
\caption{Hyperparameters of \protect\texttt{LightGBM} optimized in the
benchmark experiments. The table shows the lower and upper bounds of the
search space for each hyperparameter, along with any special properties
such as logscale
transformations.}\label{tbl-lightgbm-hyperparameters}\tabularnewline
\toprule\noalign{}
Hyperparameter & Lower & Upper & Properties \\
\midrule\noalign{}
\endfirsthead
\toprule\noalign{}
Hyperparameter & Lower & Upper & Properties \\
\midrule\noalign{}
\endhead
\bottomrule\noalign{}
\endlastfoot
\texttt{learning\_rate} & 1e-3 & 1 & Logscale \\
\texttt{feature\_fraction} & 0.1 & 1 & \\
\texttt{min\_data\_in\_leaf} & 1 & 200 & \\
\texttt{num\_leaves} & 10 & 255 & \\
\texttt{extra\_trees} & - & - & Logical value \\
\texttt{lambda\_l1} & 1e-3 & 1e3 & Logscale \\
\texttt{lambda\_l2} & 1e-3 & 1e3 & Logscale \\
\texttt{min\_gain\_to\_split} & 1e-3 & 0.1 & Logscale \\
\texttt{num\_iterations} & 1 & 5,000 & Internal tuning parameter \\
\end{longtable}

\begin{longtable}[]{@{}rlrrr@{}}
\caption{Datasets used in the benchmark experiments. The table shows the
OpenML \citep{bischl2025} task ID, name, number of instances, number of
features, and number of classes for each
dataset.}\label{tbl-datasets}\tabularnewline
\toprule\noalign{}
Task ID & Name & Instances & Features & Classes \\
\midrule\noalign{}
\endfirsthead
\toprule\noalign{}
Task ID & Name & Instances & Features & Classes \\
\midrule\noalign{}
\endhead
\bottomrule\noalign{}
\endlastfoot
31 & credit-g & 1,000 & 20 & 2 \\
3945 & KDDCup09\_appetency & 50,000 & 230 & 2 \\
7592 & adult & 48,842 & 14 & 2 \\
189354 & airlines & 539,383 & 7 & 2 \\
\end{longtable}

\begin{longtable}[]{@{}
  >{\raggedright\arraybackslash}p{(\linewidth - 14\tabcolsep) * \real{0.1473}}
  >{\raggedright\arraybackslash}p{(\linewidth - 14\tabcolsep) * \real{0.0775}}
  >{\raggedleft\arraybackslash}p{(\linewidth - 14\tabcolsep) * \real{0.1240}}
  >{\raggedleft\arraybackslash}p{(\linewidth - 14\tabcolsep) * \real{0.1163}}
  >{\raggedleft\arraybackslash}p{(\linewidth - 14\tabcolsep) * \real{0.1008}}
  >{\raggedleft\arraybackslash}p{(\linewidth - 14\tabcolsep) * \real{0.0853}}
  >{\raggedleft\arraybackslash}p{(\linewidth - 14\tabcolsep) * \real{0.1860}}
  >{\raggedleft\arraybackslash}p{(\linewidth - 14\tabcolsep) * \real{0.1628}}@{}}
\caption{Runtime breakdown of the BO algorithms across the four
benchmark tasks. For each task--algorithm combination, the table reports
the cumulative runtime of the learners, the surrogate model, and the
optimizer in seconds, the wall-clock time in seconds, the learner
utilization in percent, and the final classification error. The
learners, surrogate, and optimizer columns are cumulative runtimes
summed across all workers. All entries are the mean over five
replications with the standard deviation in
parentheses.}\label{tbl-benchmarks-extra}\tabularnewline
\toprule\noalign{}
\begin{minipage}[b]{\linewidth}\raggedright
Task
\end{minipage} & \begin{minipage}[b]{\linewidth}\raggedright
Algorithm
\end{minipage} & \begin{minipage}[b]{\linewidth}\raggedleft
Learners (s)
\end{minipage} & \begin{minipage}[b]{\linewidth}\raggedleft
Surrogate (s)
\end{minipage} & \begin{minipage}[b]{\linewidth}\raggedleft
Optimizer (s)
\end{minipage} & \begin{minipage}[b]{\linewidth}\raggedleft
Wall-clock (s)
\end{minipage} & \begin{minipage}[b]{\linewidth}\raggedleft
Learner Utilization (\%)
\end{minipage} & \begin{minipage}[b]{\linewidth}\raggedleft
Classification Error
\end{minipage} \\
\midrule\noalign{}
\endfirsthead
\toprule\noalign{}
\begin{minipage}[b]{\linewidth}\raggedright
Task
\end{minipage} & \begin{minipage}[b]{\linewidth}\raggedright
Algorithm
\end{minipage} & \begin{minipage}[b]{\linewidth}\raggedleft
Learners (s)
\end{minipage} & \begin{minipage}[b]{\linewidth}\raggedleft
Surrogate (s)
\end{minipage} & \begin{minipage}[b]{\linewidth}\raggedleft
Optimizer (s)
\end{minipage} & \begin{minipage}[b]{\linewidth}\raggedleft
Wall-clock (s)
\end{minipage} & \begin{minipage}[b]{\linewidth}\raggedleft
Learner Utilization (\%)
\end{minipage} & \begin{minipage}[b]{\linewidth}\raggedleft
Classification Error
\end{minipage} \\
\midrule\noalign{}
\endhead
\bottomrule\noalign{}
\endlastfoot
credit-g & CL & 326 (22) & 201 (13) & 331 (18) & 575 (25) & 0.1 (0.0) &
0.1918 (0.0030) \\
& ACBO & 287 (14) & 235 (4) & 356 (5) & 593 (2) & 0.1 (0.0) & 0.1910
(0.0028) \\
& ADBO & 11,164 (325) & 211,614 (413) & 29,172 (340) & 594 (1) & 4.2
(0.1) & 0.1838 (0.0016) \\
KDDCup09 appetency & CL & 3,493 (358) & 23 (1) & 74 (1) & 725 (2) & 1.1
(0.1) & 0.0175 (0.0000) \\
& ACBO & 12,408 (521) & 239 (5) & 348 (5) & 588 (2) & 4.7 (0.2) & 0.0175
(0.0000) \\
& ADBO & 201,426 (2,170) & 37,155 (1,676) & 9,165 (221) & 590 (0) & 76.2
(0.8) & 0.0174 (0.0000) \\
adult & CL & 4,271 (523) & 123 (10) & 242 (14) & 580 (36) & 1.6 (0.1) &
0.1227 (0.0001) \\
& ACBO & 5,809 (1,486) & 230 (7) & 362 (7) & 592 (2) & 2.2 (0.6) &
0.1226 (0.0003) \\
& ADBO & 149,507 (2,067) & 91,394 (2,003) & 13,774 (300) & 594 (0) &
56.2 (0.8) & 0.1222 (0.0001) \\
airlines & CL & 27,396 (7,143) & 19 (4) & 61 (12) & 664 (103) & 9.1
(1.6) & 0.3236 (0.0004) \\
& ACBO & 83,108 (20,212) & 220 (21) & 364 (18) & 585 (4) & 31.7 (7.6) &
0.3235 (0.0003) \\
& ADBO & 260,517 (929) & 293 (348) & 567 (421) & 586 (1) & 99.2 (0.5) &
0.3233 (0.0001) \\
\end{longtable}

\end{document}